\documentclass[11pt, a4paper]{article}

\usepackage[a4paper, total={16cm, 25cm}]{geometry}
\usepackage{amsmath}
\usepackage{graphicx}
\usepackage{natbib}
\usepackage{amsthm}
\usepackage{amsfonts}
\usepackage{color}
\usepackage{multirow}
\usepackage{appendix}
\usepackage{cleveref}
\usepackage{url}
\usepackage{xr}
\newtheorem{rational for conjecture}{Rational for Conjecture}

\begin{document}

\huge

\noindent \textbf{Estimating Tukey Depth Using Incremental Quantile Estimators}
%  Real-time Characterization of Multivariate Data Stream Distributions Using the Concept of Depth}

\large

\begin{center}
Hugo Lewi Hammer\footnote{Oslo Metropolitan University and Simula Metropolitan Center} \footnote{Corresponding author. Email: \texttt{hugo.hammer@oslomet.no}}, Anis Yazidi\footnote{Oslo Metropolitan University} and H{\aa}vard Rue\footnote{King Abdullah University of Science \& Technology}
\end{center}

\normalsize

\begin{abstract}
  The concept of depth represents methods to measure how deep an arbitrary point is positioned in a dataset and can be seen as the opposite of outlyingness. It has proved very useful and a wide range of methods have been developed based on the concept. 
  
  To address the well-known computational challenges associated with the depth concept, we suggest to estimate Tukey depth contours using recently developed \textit{incremental quantile estimators}. The suggested algorithm can estimate depth contours when the dataset in known in advance, but also \textit{recursively update} and even \textit{track} Tukey depth contours for dynamically varying data stream distributions. Tracking was demonstrated in a real-life data example where changes in human activity was detected in real-time from accelerometer observations.

%  The algorithms estimated Tukey depth contours equally well for both elliptic (Gaussian) and non-elliptic distributions. The performance, however, depends on the degree of curvature for the true depth contours being closely related to the degree of dependency between variables. For static data streams, the algorithm estimated a depth contour of dimension $p=10$ with a mean absolute error in Tukey depth less than 0.01 in 1.2 and 125 minutes for independent and strongly dependent variables, respectively, on a single CPU processor. We have not found any study that has been able to estimate depth contours of such a high dimension documenting the efficiency of the algorithm. For dynamically changing data streams, even for dimensions as high as $p=5$, the algorithm was able to process tens of thousands of observations per second and track depth contours with high precision.
\end{abstract}

\section{Introduction}

Several attempts have been made to provide a desirable ordering of multivariate data and the concept of depth has become very popular. Depth gives a center-outward ordering and a wide range of depth measures have been developed \citep{mosler2013depth}, such as depth based on distance metrics (Mahalanobis, spherical, projection and oja), weighted mean depths \citep{dyckerhoff2011weighted} and depth based on halfspaces and simplices \citep{tukey1975mathematics,zhang2002some,liu1999multivariate}. The concept has further been extended to measure the depth of regression models (regression depth) \citep{rousseeuw1999regression} and the depth of functional data \citep{lopez2009concept,lopez2014simplicial}.

The depth concept has been used to develop a range of new statistical models and methods for classification and clustering \citep{kim2018data,hubert2017multivariate,jornsten2004clustering}, functional autoregressive models \citep{martinez2019robust}, characterization of multivariate data \citep{serfling2004nonparametric} such as  multivariate kurtosis \citep{wang2005nonparametric}, multi quantile regression \citep{paindaveine2012computing2,paindaveine2012computing} and robust principal component analysis \citep{mozharovskyi2016tukey}. Depth has been applied to a wide range of disciplines such as economy \citep{kim2018data,kosiorowski2014depthproc,hubert2017multivariate}, health and biology \citep{williams2008modelling,hubert2015multivariate}, ecology \citep{cerdeira2018revisiting} and hydrology \citep{chebana2011depth} to name a few.

%Unfortunately, the most popular depth measures are  
%For most of the depth based methods above it is useful to compute depth for fairly high dimensions, for example to do depth based classification or clustering. Unfortunately, the most popular depth measures are computationally demanding for higher dimensions.

The earliest and most popular depth measure is Tukey depth \citep{tukey1975mathematics}. The Tukey depth of a point is defined as the minimum probability mass carried by any closed halfspace containing the point. However, computation of Tukey depth for higher dimensions is computationally demanding limiting its application \citep{liu2019fast}.

\citet{kong2012quantile} defined halfspaces such that a specific portion of observations are on one side of the halfspace. \citet{kong2012quantile} showed that contours with a specific Tukey depth can be estimated from the intersection of such halfspaces over different directions. Such contours can again be used to estimate the depth of any point. 

To apply this result to estimate $\alpha$-depth contours in dimension $p$, then the positions of $O(c^{p-1})$, $c>1$ halfspaces must be estimated requiring estimators that are both memory and computationally efficient. In this paper we suggest to use \textit{incremental quantile estimators} \citep{hammer2018new,hammer2019joint}. These estimators only need to store a single value in memory, $O(1)$, and only need to perform a single operation per observation resulting in a computational complexity of $O(n)$ for $n$ observations. In comparison, traditional quantile estimators have a memory requirement of $O(n)$ and a $O(n \log n)$ computational complexity.
Thus streaming quantile estimators are ideal to estimate high dimensional $\alpha$-depth contours.
However, the computational efficiency comes with a price and traditional estimators provide more precise estimates based on the same observations. We will demonstrate that incremental quantile estimators are useful when the data is known in advance, but the main application is for streams of data since the estimators are able to \textit{recursively} update and even \textit{track} the true positions of the halfspaces.
%The methods can even \textit{track} Tukey depth contours of dynamically varying data stream distributions.
Tracking will be demonstrated in a real-life data example where changes in human activity are detected in real-time from accelerometer observations.

The QEWA and CondQ incremental quantile estimators, in \citet{hammer2018new} and \citet{hammer2019joint}, document state-of-the-art tracking performance, but these estimators are based on generalized exponentially weighted averages and are not robust to outliers which are common in real-life data streams. Among quantile estimators being robust to outliers, the Deterministic Update Based Multiplicative Incremental Quantile Estimator (DUMIQE) in \citet{yazidi17} and the ShiftQ estimator in \citet{hammer2019joint} document state-of-the-art performance \citep{hammer2019joint}, and thus the $\alpha$-depth contour estimators in this paper will be based on these estimators. 

The paper is organized as follows. In Section \ref{sec:depth} the concept of depth is introduced including some results to compute Tukey depth. Section \ref{sec:alg} provides an efficient procedure to estimate Tukey depth. Section \ref{sec:measure} presents performance metrics that will be used to evaluate the algorithm and Sections \ref{ref:syntexp} and \ref{sec:real} present synthetic and real-life data experiments.

\section{The Concept of Depth}
\label{sec:depth}

% * Concept of depth. Forklare hvorfor ikke konturene til simultanfordelingen (lokal egenskap til fordelingen). De fire egenskapene. Eksempler pÃ¥ ulike depth mÃ¥l hvor Tukey depth presenteres fÃ¸rst og i stÃ¸rst detalj.\\
% * Directional quantiles, contour og alpha depth region

Let $X = (X_1, \ldots, X_p)^T$ represent a $p$-dimensional stochastic vector with probability distribution $P$. Let $D(x,P)$ denote the \textit{depth function} of a point $x$ with respect to the probability distribution $P$. A high (low) value of the depth function refers to a central (outlying) point of the probability distribution. A general depth function is defined by satisfying the natural requirements of affine invariance, maximality at center, monotonicity relative to deepest point and vanishing at infinity \citep{zuo2000general}.
%Several depth function have been suggested, see e.g. \citet{zuo2000general} and \citet{mosler2013depth}.

The most used and popular depth function is Tukey depth defined as the minimum probability mass carried by any closed halfspace containing the point
\begin{align}
  \label{eq:1}
  % d(x,P) = \inf_{u \in \mathcal{S}^{p-1}} P\left( \left\{z \in \mathbb{R}^p : u^Tx \leq u^Tz \right\} \right)
  D(x,P) = \inf_{u \in \mathcal{U}} P\left(u^TX \leq u^Tx \right)
\end{align}
where $\mathcal{U}$ refers to the set of all vectors with unit length.
%Tukey depth is also referred to as halfspace depth or location depth in the literature.
%Tukey depth satisfies the general requirements of a depth function as described above.

Define the $\alpha$-depth region with respect to Tukey depth, $D(\alpha)$, as the set of points whose depth is at least $\alpha$
\begin{align}
  \label{eq:2}
  D(\alpha) = \left\{x \in \mathbb{R}^p : D(x,P) \geq \alpha\right\}
\end{align}
The halfspace depth regions are closed, convex, and nested for increasing $\alpha$. The boundary of $D(\alpha)$ is known as the $\alpha$-depth contour.
%The halfspace median (or Tukey median) is defined as the center of gravity of the smallest non-empty depth region, i.e. the region containing the points with maximal Tukey depth.

Following \citet{kong2012quantile}, for any unit directional vector $u \in \mathcal{U}$, define the \textit{directional quantile} as
\begin{align}
  \label{eq:3}
  Q(\alpha,u^TX) &= F^{-1}_{u^TX}(\alpha)
\end{align}
where $F_{u^TX}(x)$ refers to the univariate cumulative distribution function of the projection of $X$ on $u$. Define the halfspace
\begin{align}
  \label{eq:4}
  H(\alpha,u) = \left\{x \in \mathbb{R}^p : u^Tx \geq Q(\alpha,u^TX)\right\}
\end{align}
which is bounded away from the origin at distance $Q(\alpha,u^TX)$ by the hyperplane with normal vector $u$. Consequently $P(X \in H(\alpha,u)) = 1-\alpha$ for any $u \in \mathcal{U}$.

\citet{kong2012quantile} proved that the $\alpha$-depth region in \eqref{eq:2} equals the \textit{directional quantile envelope} 
\begin{align}
  \label{eq:5}
  D(\alpha)= \bigcap_{u \in \mathcal{U}} H(\alpha,u)
\end{align}
Tukey depth may not be defined for depths above some threshold and the intersection becomes empty.
% In the next section an algorithm is presented that uses this result to estimate $D(\alpha)$. 
%The result in \eqref{eq:5} lay the foundation for computationally efficient algorithms to estimate $D(\alpha)$. Estimate $Q(\alpha,u)$ for a set of directional vectors $u$ and compute the resulting envelope. 

\section{Efficient Estimation of Tukey Depth}
\label{sec:alg}

For a given multivariate dataset, the result in \eqref{eq:5} invites to a simple procedure to estimate $\alpha$-depth regions. Simply select a set of directional vectors, $u_i, i = 1, \ldots, n_u$, and estimate the directional quantiles as given in \eqref{eq:3} from the dataset. To get a fully functional algorithm, four issues will be discussed below.

\begin{itemize}
\item[\textbf{1.}] \textbf{Estimating directional quantiles.} As pointed out in the introduction, we suggest to use incremental quantile estimators. 
%traditional quantile estimators are both memory and computationally demanding.
%Thus if data is received as a data stream or the dimension or amount of data is high, estimation of $\alpha$-depth regions based on traditional quantile estimators will be computationally infeasible.
%In this paper we suggest to rather use incremental quantile estimators which are far less memory and computationally demanding. %Estimates of $\alpha$-depth contours can then be obtained by computing directional quantile estimates using ShiftQ.
%Incremental quantile estimators are based on recursively and incrementally update the quantile estimates for every data point \citep{Tierney1983}. The recursive updating makes them particularly useful to estimate and even %track quantiles of dynamically varying data stream distributions.
A prominent example is the DUMIQE algorithm which can update directional quantile estimates as follows when an observation $x_n$ is received \citep{yazidi17}
\begin{align}
  \label{eq:6}
  \begin{split}
  \widehat{Q}(\alpha,u_i^TX_{n}) &\leftarrow (1 + \lambda \alpha) \widehat{Q}(\alpha,u_i^TX_{n-1}) , \hspace{11mm}\text{ if } u_i^Tx_{n} > \widehat{Q}(\alpha,u_i^TX_{n-1}) \\
  \widehat{Q}(\alpha,u_i^TX_{n}) &\leftarrow (1 - \lambda (1-\alpha)) \widehat{Q}(\alpha,u_i^TX_{n-1}) , \hspace{2mm}\text{ if } u_i^Tx_{n} < \widehat{Q}(\alpha,u_i^TX_{n-1})
  \end{split}
\end{align}
The update is quite intuitive. If the sample $u_i^Tx_n$ is above (respectively below) the current estimate, increase (respectively reduce) the corresponding directional quantile estimate. The tuning parameter $\lambda > 0$ controls the update size. For example, if the distribution of $X_n$ changes rapidly with time, a high value of $\lambda$ should be selected to efficiently track the directional quantiles. The $\alpha$-depth region can now be estimated with the intersection 
\begin{align}
  \label{eq:11}
  \widehat{D}_{n}(\alpha) = \bigcap_{i\, \in\, 1,\ldots,n_u} \widehat{H}_n(\alpha,u_i), \,\, k=1,\ldots,K
\end{align}
where halfspaces are defined from the directional quantile estimates
\begin{align}
  \label{eq:10}
  \widehat{H}_n(\alpha,u_i) = \left\{x \in \mathbb{R}^p : u_i^Tx \geq \widehat{Q}(\alpha,u_i^TX_n) \right\},
\end{align}
%$\widehat{D}_{n}(\alpha)$ can become empty, but his happened rarely in the experiments and only for very poor choices of tuning parameter values in the incremental quantile estimators.

Often it is useful with joint estimates of multiple $\alpha$-depth contours for example to efficiently estimate the depth of any point. The ShiftQ incremental algorithm makes joint quantile estimates for multiple probabilities and in particular ensures that the ordering of quantile estimates \citep{hammer2019joint}. Thus using ShiftQ, the resulting $\alpha$-depth contours will not intersect.
\item[\textbf{2.}] \textbf{Estimating depth of any point.} If multiple $\alpha$-depth contours are estimated, the depth of any point $w$ can be estimated by checking which of the halfspaces $w$ is within to find which of the $\alpha$-depth regions $w$ is within. The naive algorithm runs in $O(n_u K)$ time, but can be sped up. If $w$ is outside $\widehat{H}_n(\alpha_j,u_i)$ for some direction $u_i$ and depth $\alpha_j$, $w$ will be outside all regions with smaller depth, $\widehat{D}_{n}(\alpha_k), \,\, k=1,\ldots,j$.
%and it is not necessary to check if $w$ is within or outside the halfspaces $\widehat{H}_n(\alpha_k,u_i), k=1,\ldots,j$ for any of the other directions.
By checking halfspaces for decreasing depth will substantially speed up the computations. 
\item[\textbf{3.}] \textbf{Selecting directional vectors.} Generation of uniformly distributed directional vectors is simple: Let $Z_1, \ldots, Z_p$ be independent standard normally distributed stochastic variables and define $Z = (Z_1, \ldots, Z_p)^T$. Then $U = Z/\|Z\|_2$ will be uniformly distributed on the unit sphere, where $\| \cdot \|_2$ refers to the Euclidean norm. Intuitively, it makes sense to use directional vectors that are more equidistantly spread on the unit sphere. A simple approach is to generate many uniformly distributed directional vectors, $N_u$, and secondly filter out directional vectors that are closer than some threshold. The approach is however computationally demanding, $O(N_u^2p^2)$, but only needs to be done in the initialization of the algorithm.
%For static datasets and for shorter dynamical data streams it is usually better, with respect to total computation time, to use uniformly distributed directional vectors. Generation of more equidistantly spread direction %vectors just takes too much time.
There are other algorithms to generate fairly equidistantly spread direction vectors, see e.g spiral algorithms \citep{saff1997distributing}. We have not evaluated the potential of these algorithms.
The optimal distribution of direction vectors also depends on the shape of the multivariate distribution, e.g. directions where the $\alpha$-depth contours have strong curvature, more directional vectors should be used. One can imagine to recursively update the directional vectors as one learns more about the distribution. We have not looked into this.
\item[\textbf{4.}] \textbf{Convergence.} $\widehat{D}_{n}(\alpha)$ consists of two approximations compared to $D(\alpha)$ namely the finite number of directional vectors and the quantile estimates. Thus for $\widehat{D}_{n}(\alpha)$ to converge to $D(\alpha)$, first, the directional vector selection procedure must cover the unit sphere when the number of directional vectors goes to infinity and, secondly, the directional quantile estimates must converge to the true directional quantiles, when the number of observations goes to infinity. By using the simple procedure above to select uniformly distributed directional vectors, the first requirement is satisfied. Further \citet{yazidi17} and \citet{hammer2019joint} prove the second requirement.
  %Thus $\widehat{D}_{n}(\alpha)$ converges to $D(\alpha)$ with an increasing number of observations and using an increasing number of directional vectors.
\end{itemize}

\section{Performance Metrics}
\label{sec:measure}

\begin{figure}
  \centering
  \includegraphics[width = 0.7\textwidth]{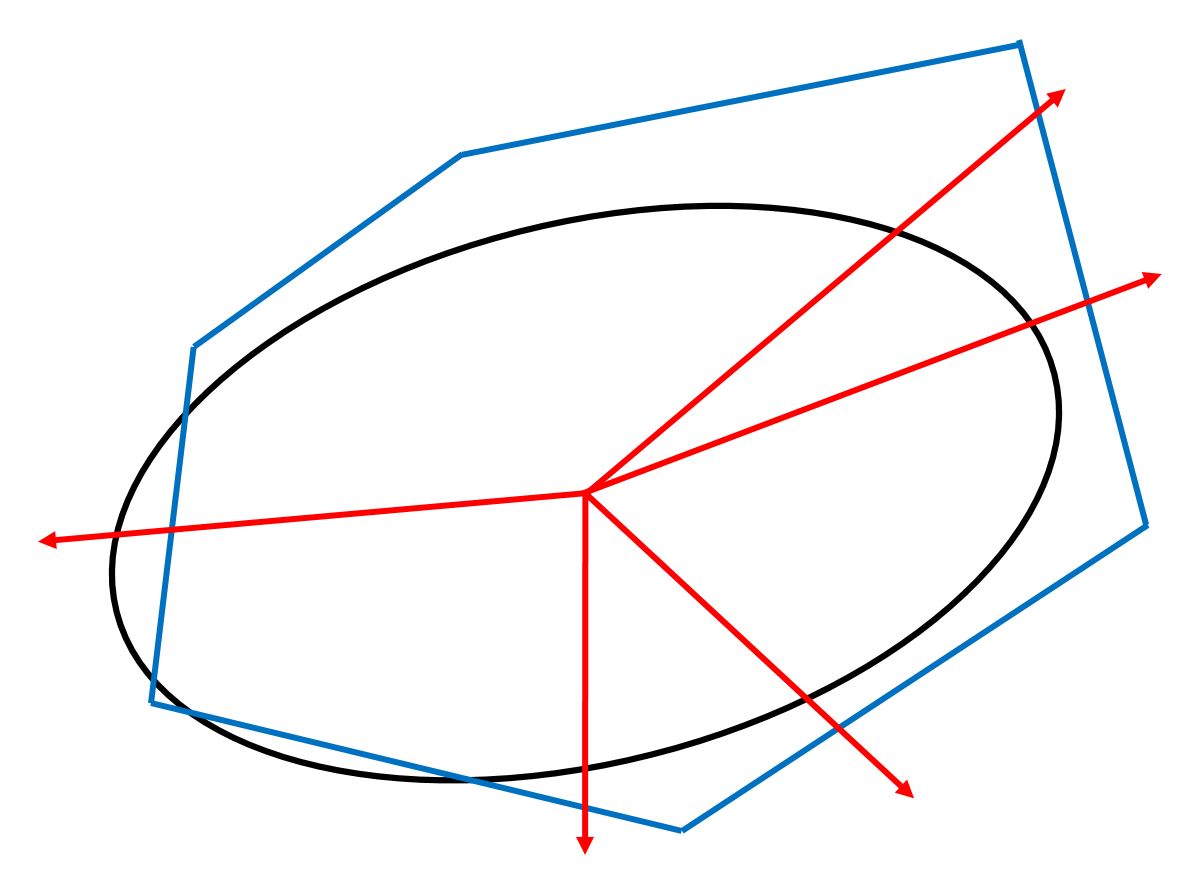}
  \caption{The approach to measure $\alpha$-depth contour estimation error. The black and blue curves show the true $\alpha$-depth regions and the envelope estimate. The lines with directions $v_i, i = 1, \ldots, n_v$ are shown in red.}
  \label{fig:1}
\end{figure}

% It is not obvious how to measure the error between a true $\alpha$-depth region, $D(\alpha_k)$, and an estimate, $\widehat{D}(\alpha_k)$.
%To evaluate the efficiency of the algorithm, reliable performance measures are important.
We suggest to measure error along lines $l_i, i = 1, \ldots, n_v$ going trough the center of the true distribution and outward in uniformly distributed directions $v_i, i = 1, \ldots, n_v$ (Figure \ref{fig:1}). This approach scales well with dimension $p$. 
We suggest two error measures:
\begin{itemize}
\item[\textbf{1.}] \textbf{Depth error:} Let $\widetilde{w}_{i,k}$ denote the point of intercept between the line, $l_i$, and the envelope and compute the true depth at this point, $D(\widetilde{w}_{i,k}, P)$. The error is computed using mean absolute depth error (MADE) over all the lines $l_i, i=1,\ldots, n_v$
  \begin{align*}
    \text{MADE}_k = \frac{1}{n_v} \sum_{i=1}^{n_v} \left| \alpha_k - D(\widetilde{w}_{i,k}, P) \right|
  \end{align*}
and again average over envelopes
\begin{align}
    \label{eq:15}
  \text{MADE} = \frac{1}{K} \sum_{k=1}^K \text{MADE}_k
\end{align}
To compute MADE for higher dimensions, the true depth must to be computed for a large set of points $\widetilde{w}_{i,k}$. For non-elliptic distributions this is computationally demanding and was limited to $p \leq 6$ in the experiments. For elliptic distributions, and in particular multivariate normal distributions, the true depth of any point can be computed analytically and thus MADE was computed up to dimension $p=10$ in the experiments. Details are given in Appendix \ref{app:de}. Of course, if we knew that the observations were multivariate normally distributed, other depth measures such as Mahalanobis depth would be more natural, but the computations are only used to evaluate the performance of the algorithm for high dimensions.
\item[\textbf{2.}] \textbf{Euclidean distance:} Along each line, $l_i$, compute the point of intercept between the line and the true $\alpha$-depth contour of depth $\alpha_k$, denoted $w_{i,k}$. Compute the error as the average Euclidean distance (ED)
  \begin{align*}
    \text{ED}_k = \frac{1}{n_v} \sum_{i=1}^{n_v} \|w_{i,k} - \widetilde{w}_{i,k}\|_2
  \end{align*}
where $\widetilde{w}_{i,k}$ still refers to the intercept between line $l_i$ and the envelope. Further take average over envelopes
\begin{align}
    \label{eq:16}
  \text{ED} = \frac{1}{K} \sum_{k=1}^K \text{ED}_k
\end{align}
%The intersection in \eqref{eq:11} may become empty and MADE and ED cannot be computed.
%The convergence of the envelope to the true $\alpha$-depth region with increasing number of halfspaces, is a very nice theoretical result and the most obvious depth region estimator is the intersection of the estimated halfplanes. However, it becomes with the risk of ending up with empty depth region estimates. The halfspace estimates contain a lot of information in addition to the resulting envelope and an interesting question is if this information can be used to develop better depth region estimates. We have not looked into this in this paper, but is an interesting ally for future research. 
\end{itemize}

\section{Synthetic Experiments}
\label{ref:syntexp}

In this section the performance of the algorithm in Section \ref{sec:alg} is evaluated in several synthetic experiments. The experiments focus on streaming data, except in Section \ref{sec:syntex}.
% where data is assumed to be known in advance.
In Section \ref{sec:real} the algorithm is demonstrated in a real-life data example.

All computations where run on a Dell PowerEdge R815 server with 64 1.8 GHz AMD CPU processors and Linux Ubuntu operating system version 16.04. The experiments were implemented in R \citep{R2018}, but with the most computer intensive parts in C++ integrated using Rcpp \citep{Eddelbuettel2011,Eddelbuettel2013}. 

\subsection{Synthetic Experiments - Static Data Stream}
\label{sec:syntstat}

\begin{figure}
  \centering
    \includegraphics[width = 0.9\textwidth]{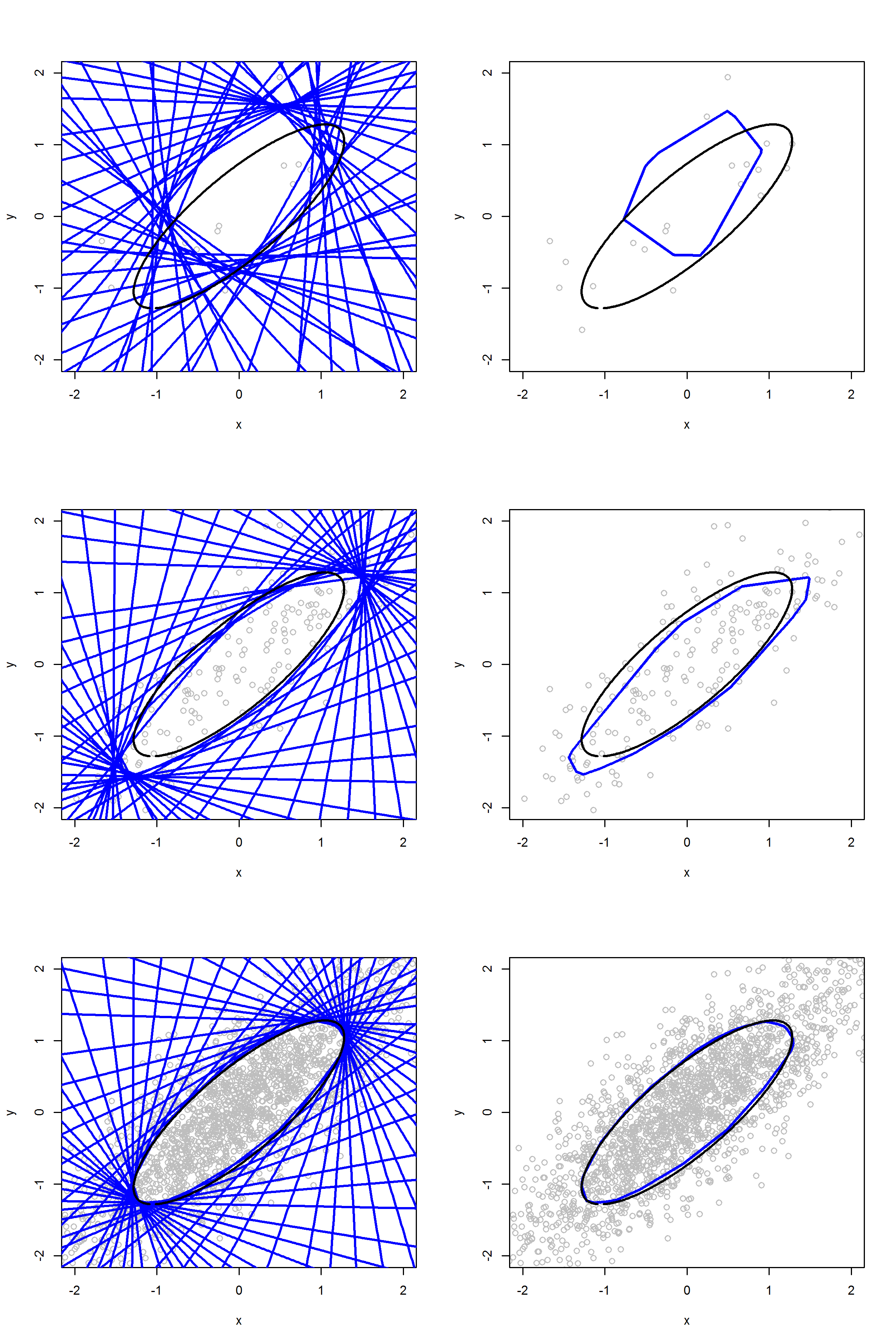}
  \caption{Multivariate normal distribution case. Estimation of $\alpha$-depth region for $\alpha = 0.1$ using $n_u = 50$ directional vectors. The rows from top to bottom show the estimates for 20, 200 and 2000 observations. The left and right column show all the half planes and the resulting envelopes in blue, respectively. The black curves show the true $\alpha$-depth contour.}
  \label{fig:3}
\end{figure}

Figures \ref{fig:3} show results of estimating the $\alpha = 0.1$ depth contour for a multivariate normally distributed data stream with parameters
\begin{align}
  \label{eq:1_2}
  \mu =
  \begin{bmatrix}
    0 \\
    0 
  \end{bmatrix}
  , \,\,
  \Sigma =
  \begin{bmatrix}
    1 & 0.82 \\
    0.82 & 1
  \end{bmatrix}
\end{align}
Directional quantiles were estimated using DUMIQE with decreasing values of the tuning parameter, $\lambda_n = 1/n$.
%\footnote{Given the results in \citet{hammer2018new} and \citet{hammer2019joint} one may expect QEWA, ShiftQ and CondQ to perform better than DUMIQE and ShiftQ. However, these papers focused on dynamic data streams and for static data streams, as considered in this section, it turned out that the DUMIQE and ShiftQ performed about equally well as QEWA, ShiftQ and CondQ. Given the simplicity of the DUMIQE and ShiftQ algorithms, the results for static data streams in the paper are based on these algorithms.}.
%The DUMIQE was initialized in the different directions using the first observation, i.e. $\widehat{Q}(\alpha,u_i^TX_1) = u_i^Tx_1$. 
%\begin{figure}
%  \centering
%    \includegraphics[width = 0.9\textwidth]{Figures/Static_nu_10}
%  \caption{Multivariate normal distribution case. Estimation of $\alpha$-depth region for $\alpha = 0.1$ using $n_u = 10$ directional vectors. The rows from top to bottom show the estimate when 20, 200 and 2000 %observations were received from the data stream. The left and right column show all the half planes and the resulting envelopes in blue, respectively. The black curves show the true $\alpha$-depth contour.}
%  \label{fig:2}
%\end{figure}
We see that fairly good estimate is achieved with 200 observations and that the error is minimal with 2000 observations. 

\begin{figure}
  \centering
    \includegraphics[width = 0.9\textwidth]{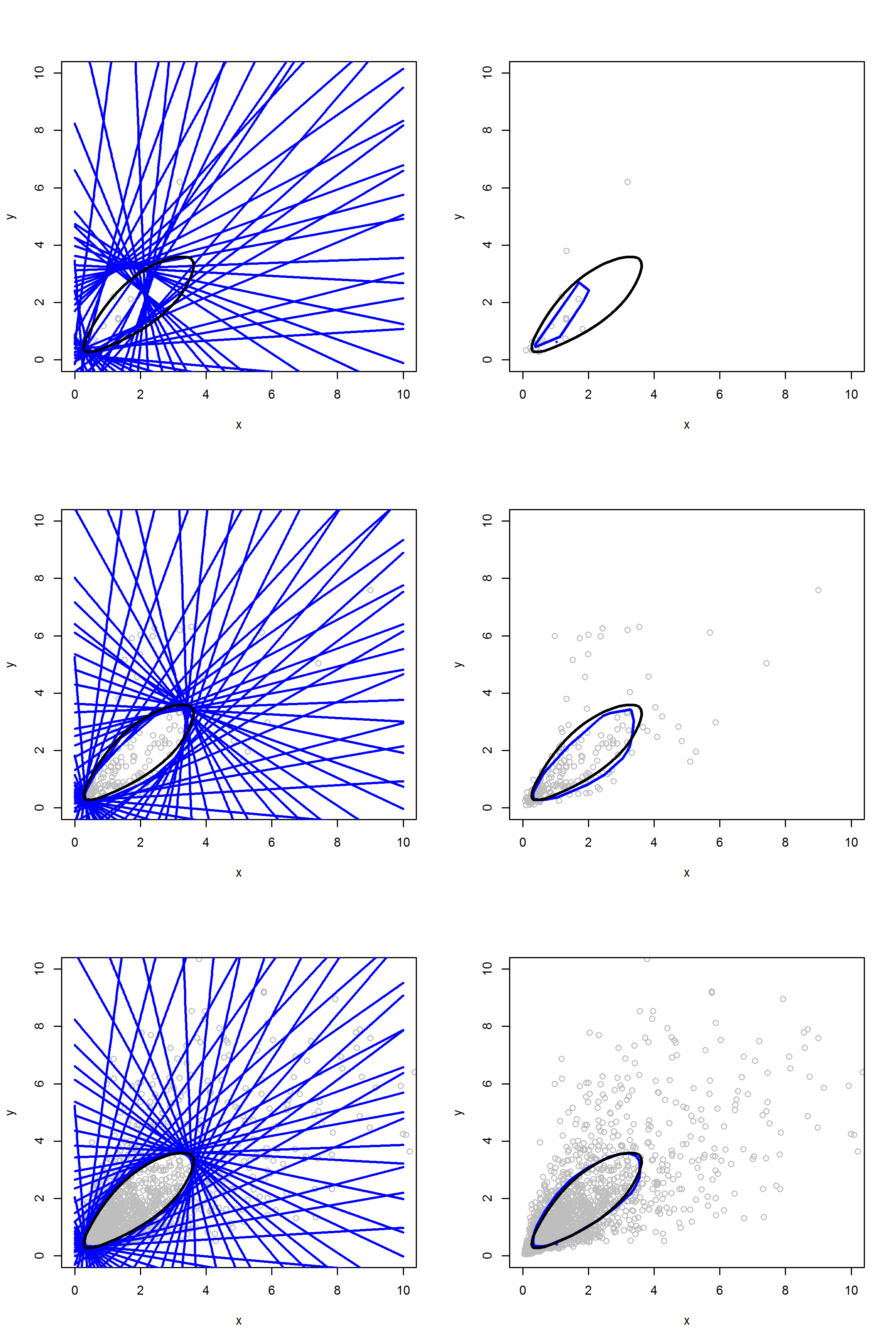}
  \caption{Multivariate lognormal distribution case. Estimation of $\alpha$-depth region for $\alpha = 0.1$ using $n_u = 50$ directional vectors. The rows from top to bottom show estimates for 20, 200 and 2000 observations. The left and right column show all the half planes and the resulting envelopes in blue, respectively. The black curves show the true $\alpha$-depth contour.}
  \label{fig:5}
\end{figure}

Let $X$ refer to the multivariate normal distribution defined above and define the lognormal distribution $Y = \exp(X)$. $Y$ is both highly unsymmetrical and highly heavy tailed. The results are shown in Figure \ref{fig:5}.
%\begin{figure}
%  \centering
%    \includegraphics[width = 0.9\textwidth]{Figures/Static_exp_nu_10}
%  \caption{Multivariate lognormal distribution case. Estimation of $\alpha$-depth region for $\alpha = 0.1$ using $n_u = 10$ directional vectors. The rows from top to bottom show the estimate when 20, 200 and 2000 %observations were received from the data stream. The left and right column show all the half planes and the resulting envelopes in blue, respectively. The black curves show the true $\alpha$-depth contour.}
%  \label{fig:4}
%\end{figure}
Due to the flexibility of the depth concept, the method performs equally well for non-elliptic distributions. 

In supplementary material S.1 a few examples of joint estimation of multiple $\alpha$-depth regions using the ShiftQ algorithm are shown. A link to the supplementary material is found at the end of the document. The results show that multiple depth regions can efficiently be estimated for both Gaussian and non-Gaussian distributions.

%The figures above give an impression of the potential of estimating $\alpha$-depth regions for streaming data.
Considered now joint estimation of $\alpha$-depth regions for $\alpha = 0.05, 0.2$ and $0.4$ and for $p > 2$.
%and the focus was on the computational time needed to obtain $\alpha$-depth region estimates with a certain amount of precision. The results were obtained using the DUMIQE algorithm with decreasing $\lambda_n = 1/n$.
%\footnote{This will not give joint estimates, but for static cases each region usually is estimated with high precision and monotone violations are unlikely. We also experimented with the ShiftQ algorithm, but for the static cases DUMIQE performed slightly better in terms of CPU time to obtain a given precision.}. The results below are based on an average of 60 independent runs and using a total of $n_v = 4p$ lines to compute MADE and ED in \eqref{eq:15} and \eqref{eq:16} essentially removing Monte Carlo error from the results. 
Table \ref{tab:4} shows results for standard multivariate normally distributed observation. More detailed results are given in Figures 5 and 6 in supplementary material S.1. A link to the supplementary material is found at the end of the document. CPU time refers to the computational time needed per $\alpha$-depth region to obtain estimates with a given precision using a single CPU core. 
      \begin{table}
        \centering
        \begin{tabular}{lcccccc}
          \hline
          & \multicolumn{2}{c}{MADE $< 0.05$} & \multicolumn{2}{c}{MADE $< 0.02$} & \multicolumn{2}{c}{MADE $< 0.01$} \\ \hline
          & CPU time & $n_u$ & CPU time & $n_u$ & CPU time & $n_u$ \\ \hline
$p=2 $      &         0.00013       &              8      &         0.00174       &             12     &          0.00942         &           18\\
$p=3 $      &         0.00023       &             12      &         0.00429       &             27     &          0.04488         &           40\\
$p=4 $      &         0.00023       &             16      &         0.00958       &             81     &          0.11631         &          122\\
$p=5 $      &         0.00043       &             20      &         0.02991       &            153     &          0.35146         &          345\\
$p=6 $      &         0.00054       &             24      &         0.07419       &            274     &          0.90636         &         1386\\
$p=8 $      &         0.00334       &             72      &         0.34695       &           1228     &          9.83845         &         9324\\
$p=10$      &         0.01361       &             90      &         1.54246       &           3450     &        104.45206         &        88412\\ \hline
        \end{tabular}
        \caption{Multivariate standard normal distribution case: The second and third columns show the CPU time (in seconds) and the number of directional vectors used to obtain MADE less than 0.05. The other columns show the same to obtain MADE less than 0.02 and 0.01, respectively.}
        \label{tab:4}
      \end{table}
The number of directional vectors (and thus CPU time) increases with $p$ and estimation precision. The algorithm performs very well. For example for dimension $p = 10$ a MADE less than 0.02 this is obtained in about 1.5 seconds. MADE $< 0.01$ could be reached in shorter CPU time than what is shown in Table \ref{tab:4} using a higher number of directional vectors, but is not explored.

Now assume that $X = (X_1, \ldots, X_p)^T$ is a multivariate normally distributed variable with zero expectation vector and strong dependencies
\begin{align}
  \label{eq:17}
\text{Cov}(X_i,X_j) = \exp(-0.2 |i-j|),\,\, i,j = 1,\ldots,p  
\end{align}
The results are shown in Table \ref{tab:5}. More detailed results are given in Figures 9 and 10 in supplementary material S.1. By comparing Tables \ref{tab:4} and \ref{tab:5}, we  see that the number of directional vectors and CPU time needed increase when the variables of $X$ are dependent.
% An interesting idea to speed up the algorithm for strongly dependent data, is to transform to more independent data, and track these transformed data and transform back.
%We will discuss this further in the closing remarks of the paper.
      \begin{table}
        \centering
        \begin{tabular}{lcccccc}
          \hline
          & \multicolumn{2}{c}{MADE $< 0.05$} & \multicolumn{2}{c}{MADE $< 0.02$} & \multicolumn{2}{c}{MADE $< 0.01$} \\ \hline
          & CPU time & $n_u$ & CPU time & $n_u$ & CPU time & $n_u$ \\ \hline
$p=2 $      &          0.00034       &             18       &        0.00622       &             40      &         0.03734       &             40 \\
$p=3 $      &          0.00095       &             27       &        0.03003       &            135      &         1.38903       &            135 \\
$p=4 $      &          0.00238       &             54       &        0.13145       &            274      &         7.47343       &            616 \\
$p=5 $      &          0.01275       &            102       &        0.43698       &            777      &         8.47334       &           3936 \\
$p=6 $      &          0.03603       &            183       &        1.81652       &           3118      &        45.88106       &          15786 \\
$p=8 $      &          0.17285       &            819       &       23.21460       &          20979      &       988.12085       &         358438 \\
$p=10$      &          0.68053       &           2300       &      245.91893       &         198927      &             -         &            -   \\ \hline
        \end{tabular}
        \caption{Multivariate normal distribution case: The second and third columns show the CPU time (in seconds) and the number of directional vectors used to obtain an mean absolute depth error (MADE) less than 0.05, respectively. The fourth and fifth and the sixth and seventh columns show the same to obtain MADE less than 0.02 and 0.01, respectively.}
        \label{tab:5}
      \end{table}

Let $X$ still represent the multivariate normally distributed variable with covariances \eqref{eq:17}.
Table \ref{tab:6} shows results for the multivariate lognormal distribution $Y = \exp(X)$. More detailed results are given in Figures 11 and 12 in supplementary material S.2. 
\begin{table}
        \centering
        \begin{tabular}{lcccccc}
          \hline
          & \multicolumn{2}{c}{MADE $< 0.05$} & \multicolumn{2}{c}{MADE $< 0.02$} & \multicolumn{2}{c}{MADE $< 0.01$} \\ \hline
          & CPU time & $n_u$ & CPU time & $n_u$ & CPU time & $n_u$ \\ \hline
$p=2 $      &      0.00013           &         8          &      0.00957        &          27        &    0.11169           &        40       \\
$p=3 $      &      0.00024           &        27          &      0.01533        &         135        &    0.56418           &       202       \\
$p=4 $      &      0.00021           &        24          &      0.03214        &         274        &    1.64312           &       924       \\
$p=5 $      &      0.00043           &        45          &      0.14592        &        1166        &    6.16044           &      3936       \\
$p=6 $      &      0.00053           &        54          &      0.27431        &        2079        &    9.30407           &     15786       \\ \hline
        \end{tabular}
        \caption{Multivariate lognormal distribution case: The second and third columns shows the CPU time (in seconds) and the number of directional vectors used to obtain an mean absolute depth error (MADE) less than 0.05, respectively. The fourth and fifth and the sixth and seventh columns show the same to obtain MADE less than 0.02 and 0.01, respectively.}
        \label{tab:6}
\end{table}
%Recall from Section \ref{sec:measure} that it is computationally demanding to estimate MADE for non-elliptic distributions and the results are limited to $p \leq 6$.
Tables \ref{tab:5} and \ref{tab:6} show that a specific level of MADE is reached faster for the lognormal distribution than for the multivariate distribution documenting that the procedure efficiently can characterize non-Gaussian distributions.

%To the best of our knowledge, the algorithm by \citet{liu2019fast} is the most efficient in the literature to estimate Tukey $\alpha$-depth regions. The authors focus on estimating exact trimmed $\alpha$-depth regions resulting in complex combinatorial algorithms and the computation burden explodes with the number of samples. In comparison, the computational burden of our algorithm increases linearly with the number of samples. The authors can document estimation results up to dimension $p=9$, but only when the number of samples are restricted to less than 80. The algorithm by \citet{liu2019fast} is not constructed to handle streaming data. 

\subsection{Synthetic Experiments - Offline Setting}
\label{sec:syntex}

In this section, we compare the performance of the incremental quantile estimator, DUMIQE, with state-of-the-art offline quantile estimators to estimate $\alpha$-depth regions when data is known in advance. State-of-the-art offline quantile estimators are based on using weighted averages of consecutive order statistics 
\begin{align*}
  Q(\alpha) = (1 - \delta) y[j] + \delta y[j+1]
\end{align*}
where $\frac{j-m}{N} \leq \alpha < \frac{j-m+1}{N}$, $y[j]$ is the $j$th order statistic of the sample, $m$ a constant and $N$ the sample size. We use $m = \frac{\alpha+1}{3}$ and $\delta = N\alpha + m - j$ and define $\alpha[k] = \frac{k - 1/3}{N + 1/3}$. The sample quantiles can be read from a linear interpolation between the points $(\alpha[k],y[k]), k=1,\ldots,N$. This the method referred to as Type 8 in the \texttt{quantile} function in R and is the one recommended by \citet{hyndman1996sample}.

We consider the multivariate normal distribution case with covariance matrix as given in \eqref{eq:17}, sample sizes $N = 500$, $2000$, $10^4$ and $5\cdot 10^4$ and dimensions $p=2$ and $p=3$. For $p=2$ and $p=3$, we used 1500 and 7500 directional vectors, respectively, which were sufficiently many for the estimation error essentially to be due to the performance of the quantile estimators.

\begin{table}
  \centering
  \begin{tabular}{clcccccc}
    \hline
    & & \multicolumn{3}{c}{$p=2$}  & \multicolumn{3}{c}{$p=3$}\\\hline
    $N$ & Method & MADE & ED & CPU & MADE & ED & CPU \\ \hline
    \multirow{2}{*}{$500$} & Offline           & 14.9  & 43.1 & 0.291&  16.9 & 40.5   &1.634 \\
                               & DUMIQE           & 25.1  & 63.9 & 0.045& 34.9  & 69.7   &0.288 \\\hline
    \multirow{2}{*}{$2000$} & Offline          & 7.0   & 20.7 & 1.421&  7.2  & 18.2   &9.489 \\
                                & DUMIQE           & 10.6 & 28.5 & 0.182& 12.2  & 26.5   &1.154 \\\hline
    \multirow{2}{*}{$10^4$} & Offline          &  3.0  & 9.0  & 8.761& 3.0   & 7.7    &55.12 \\
                                & DUMIQE           & 4.4  & 12.1 & 0.908& 4.6   & 10.6   &5.769 \\\hline
    \multirow{2}{*}{$5\cdot 10^4$} & Offline   &  1.3  & 4.0  & 52.32& 1.3   & 3.5    &326.0 \\
                                       & DUMIQE    & 1.8  & 5.4  & 4.542& 2.0   & 4.7    &28.84 \\\hline
  \end{tabular}
  \caption{Offline experiment: Comparison of the DUMIQE estimator and the estimator recommended in \citet{hyndman1996sample} to estimate $\alpha$-depth contours for $\alpha = 0.05$, 0.2 and 0.4. MADE, ED and CPU refers to the error measures  in \eqref{eq:15} and \eqref{eq:16} (multiplied by $10^3$) and CPU time used (in seconds), respectively. $N$ refers to the sample size.}
  \label{tab:1}
\end{table}
The results are shown in Table \ref{tab:1}. We see that the estimation errors using DUMIQE are about 1.5 time that of the offline estimator. If fewer directional vectors were used, the differences in estimation error were substantially reduced.
%The computational time of DUMIQE and the offline estimators are of order $O(N)$ and $O(N \ln N)$, respectively, resulting in substantial additional computation cost using the offline estimator.
Further, the computational time of the offline estimator are about ten times that of the DUMIQE estimator. In other words, if computational time or memory usage are not an issue, the offline estimator combined with a large amount of directional vectors will give the most precise estimates from the samples. Else incremental quantile estimators are preferable even for offline settings.

\subsection{Synthetic Experiments - Dynamically Changing Data Streams}

\begin{figure}
  \centering
  \begin{tabular}{cc}
    \includegraphics[width = 0.49\textwidth]{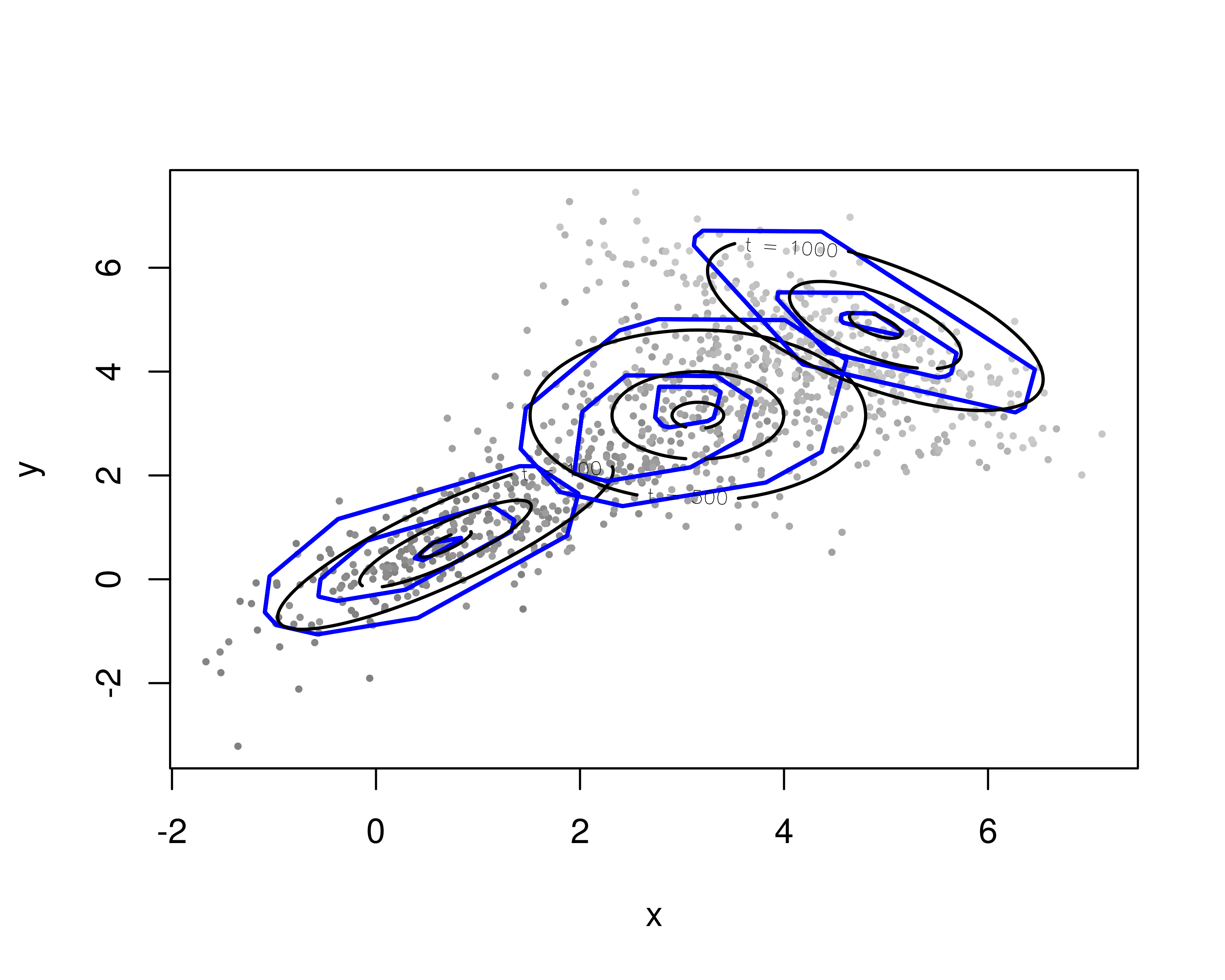} & \includegraphics[width = 0.49\textwidth]{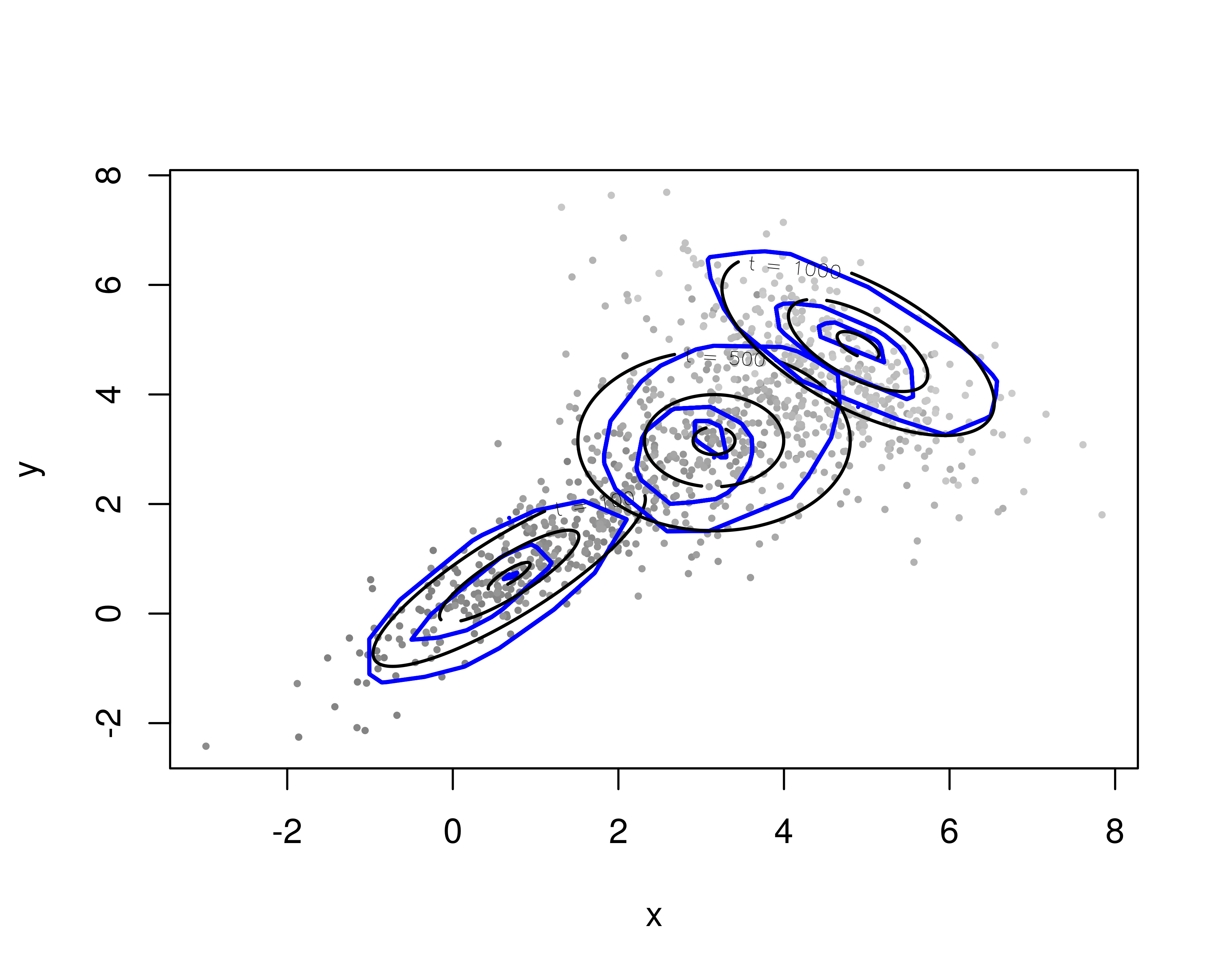} \\
    \includegraphics[width = 0.49\textwidth]{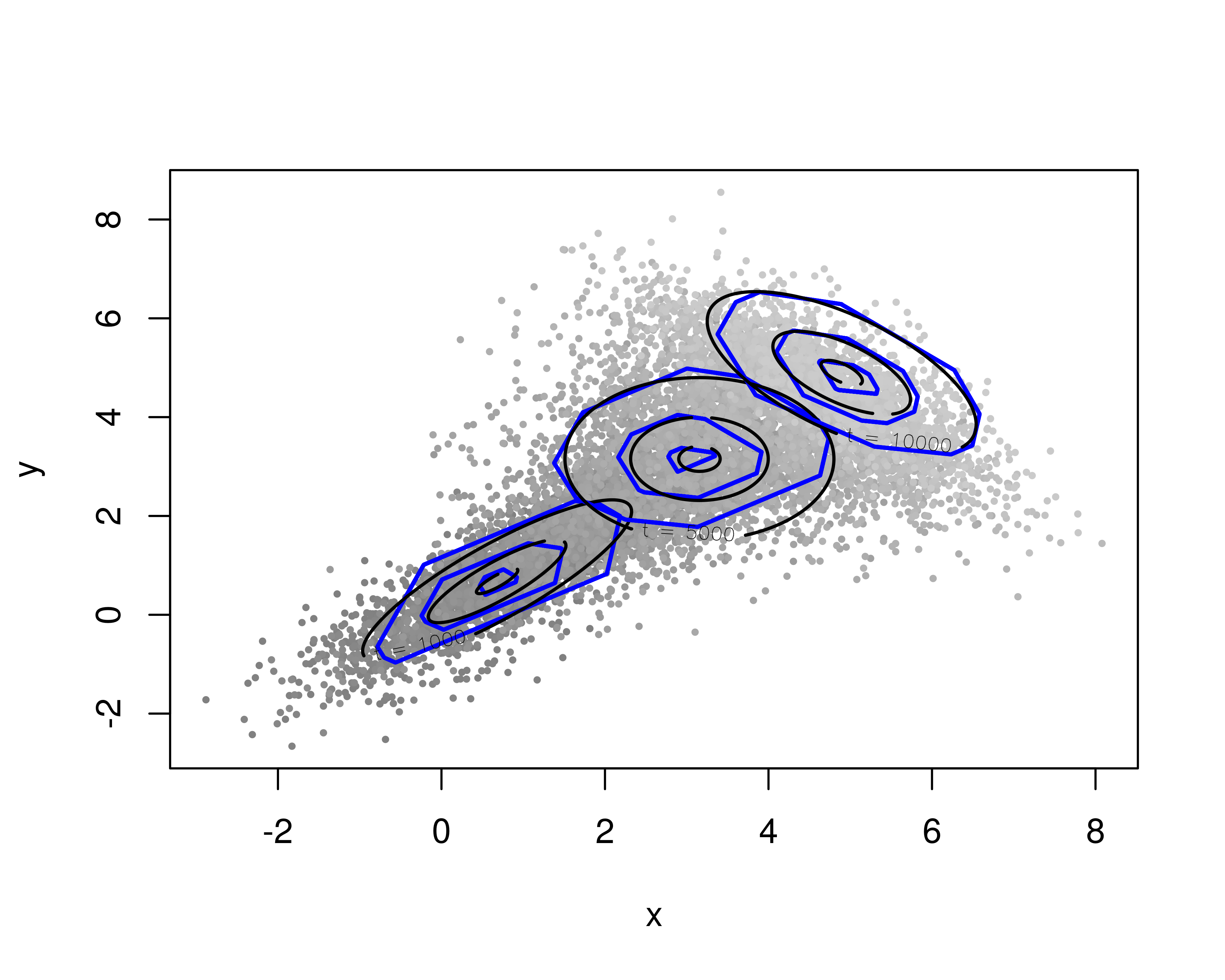} & \includegraphics[width = 0.49\textwidth]{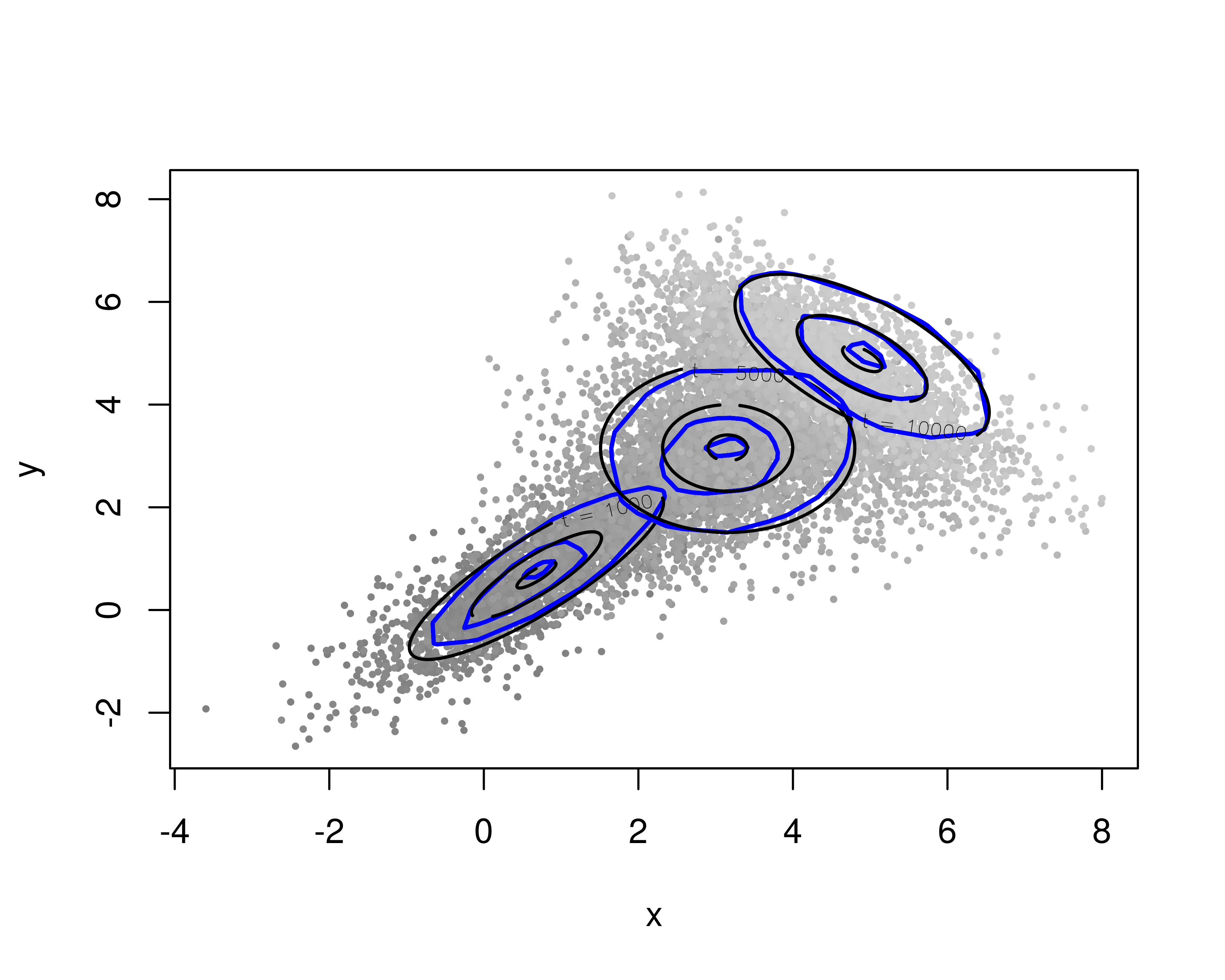} \\
    \includegraphics[width = 0.49\textwidth]{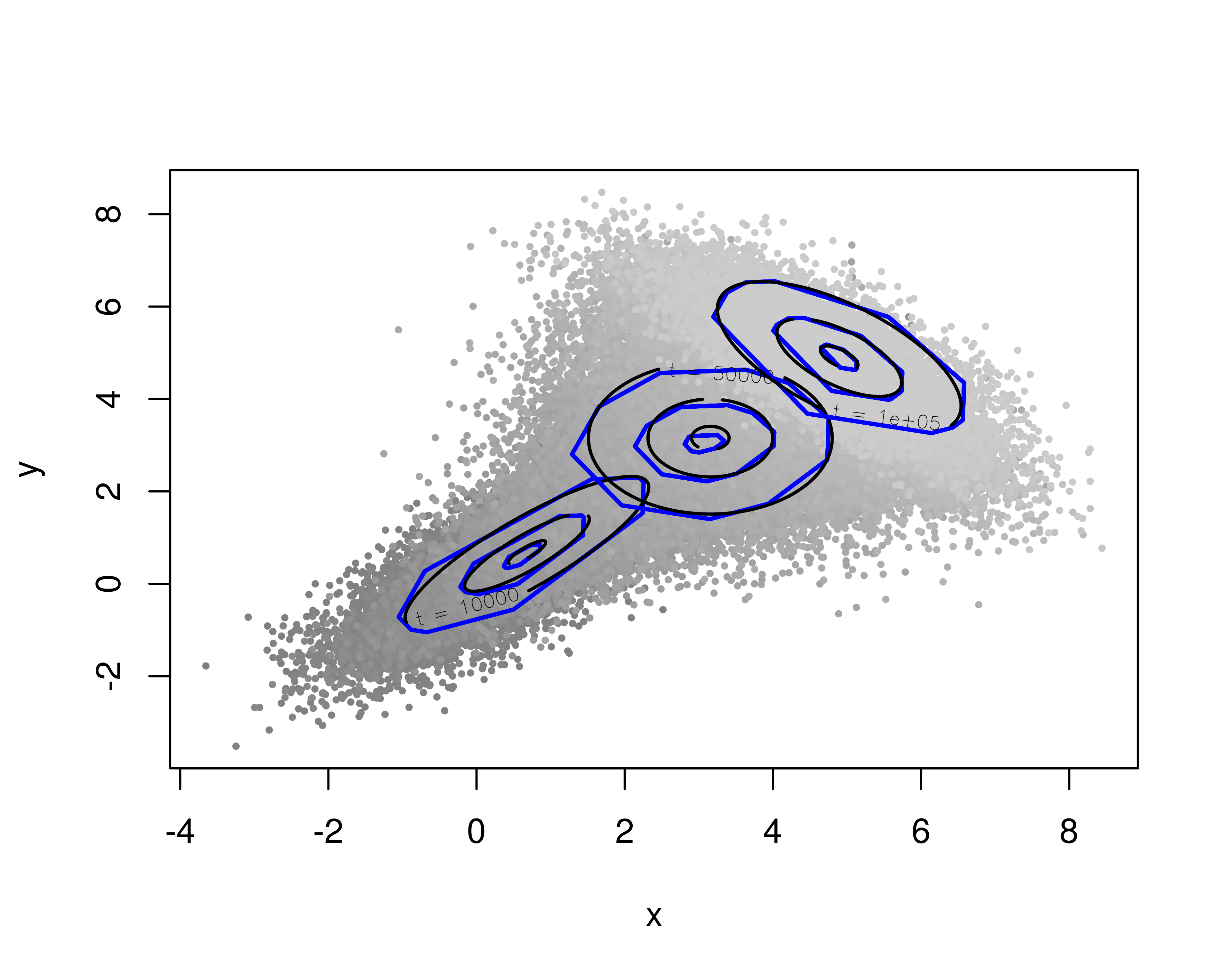} & \includegraphics[width = 0.49\textwidth]{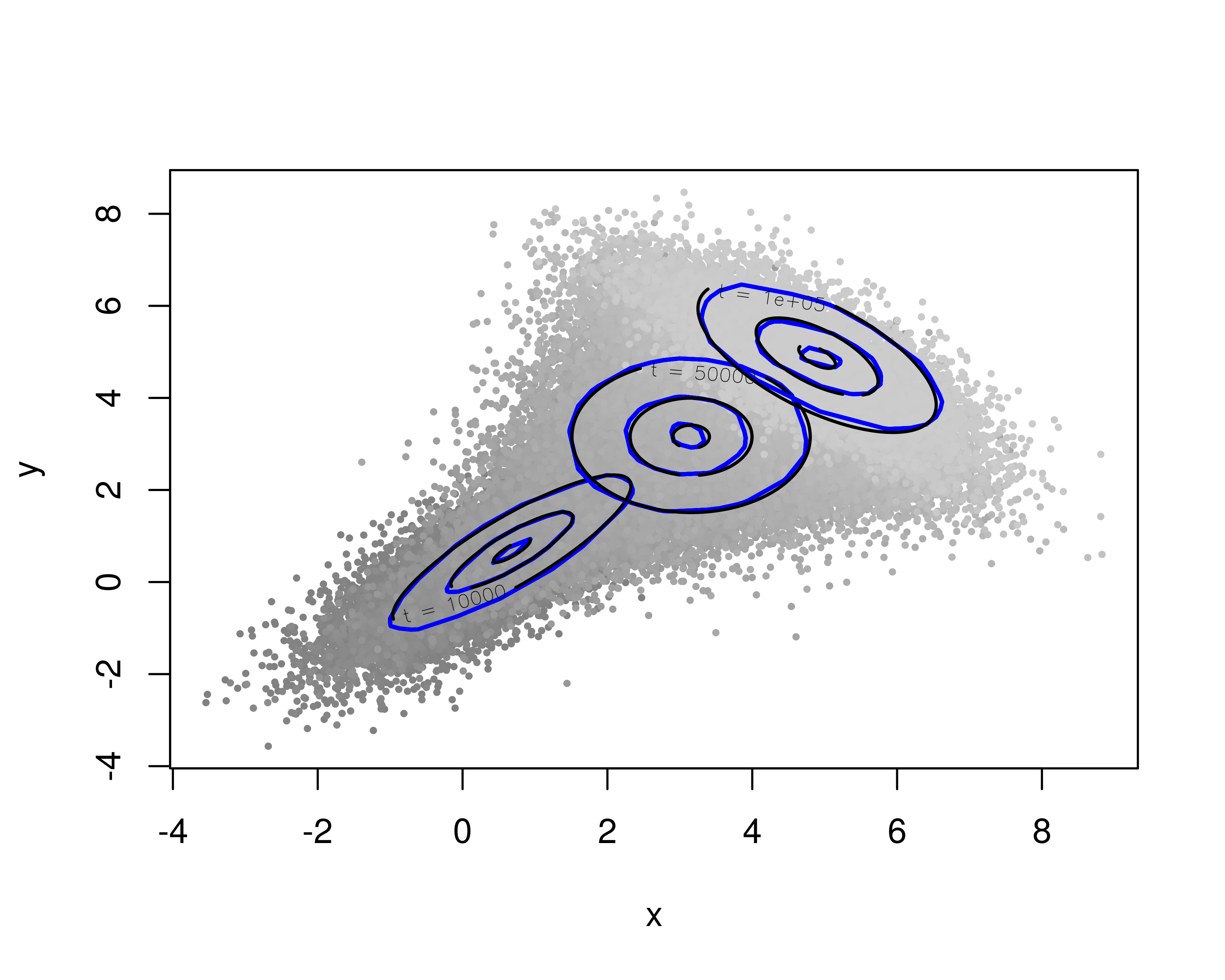} 
  \end{tabular}
    \caption{Tracking of $\alpha$-depth contours for $\alpha = 0.05$, 0.2 and 0.4: In each panel the gray dots are outcomes from the data stream. The first observations from the data stream are shown in dark gray and the dots become lighter gray as time progresses. The left and right column show cases with $n_u = 10$ and 50 directional vectors, respectively. The rows from top to bottom show cases with a total for $10^3$, $10^4$ and $10^5$ observations, respectively.}
  \label{fig:16}
\end{figure}

In this section we consider the problem of tracking $\alpha$-depth regions of dynamically varying data streams. Figure \ref{fig:16} illustrates the problem.
In each panel, the expectation vector of the data stream distribution moved from the bottom left of to the upper right. At the same time the correlation, changed from strongly positive, 0.8, to strongly negative, $-0.8$. For the $10^3$ samples case (first row), the algorithm was able to track the $\alpha$-depth regions satisfactory.
%The limited number of samples results in a fair amount of uncertainty in the directional quantile estimates and using 50 directional vectors do not perform better than using 10 directional vectors.
With $10^4$ samples the estimates improve significantly and with $10^5$ observations, the estimates are very close to the true contours. With $10^4$ and $10^5$ samples, 50 directional vectors give better and smoother estimates than 10 directional vectors. 

Evaluation for $p > 2$ is given below. Due to the computational burden of evaluating estimation error of non-elliptic distributions, the analysis was restricted to Gaussian distributions.
%The results from the previous section documented that the algorithm performed equally well for elliptic and non-elliptic distributions, and we expect the results below to be representative also for non-elliptic distributions.
Let $X_n = (X_{n,1}, \ldots, X_{n,p})^T$ be multivariate normally distributed with 
\begin{align}
  \label{eq:18}
  \mu_{n,i} = E(X_{n,i}) = \sin\left( \frac{2\pi}{T} n + \psi_i \right), \,\,\, i=1,\ldots,p
\end{align}
where $\psi_i,\,\, i=1,\ldots,p$ are independent uniformly distributed variables on the interval $[0, 2\pi]$ ensuring that the marginal expectations are out of phase. Covariance between $X_{n,i}$ and $X_{n,j}$ is
\begin{align}
  \label{eq:19}
  \text{Cov}(X_{n,i}, X_{n,j}) = \left(0.4\sin\left( \frac{2\pi}{T} n + \psi \right) + 0.4\right)^{|i-j|}
\end{align}
where $\psi$ is uniformly distributed on the interval $[0,2\pi]$.
%The $\text{Var}(X_{i,n}) = 1$ for all the variables for all the iterations and the covariance/correlation between neighbouring points varies between 0 and 0.8 for different iterations. Further, the correlation decreases exponentially with increasing $|i-j|$.

%With outcomes from the data stream with distribution as described above, we track the $\alpha$-depth regions for $\alpha = 0.05$, 0.2 and 0.4. In the ShiftQ quantile tracking algorithm we used $\lambda = \gamma$ which is in accordance with the recommendations in \citet{hammer2019joint}. To remove Monte Carlo error, for every choice of the tuning parameters, a total of $5 \cdot 10^3$ data streams of length $5 \cdot 10^4$ were run. For each new data stream, new directional vectors $u_i, i = 1, \ldots, n_u$, lines $l_i, i=1, \ldots, n_v$ and $\psi, \psi_1, \ldots, \psi_p$ were generated. To remove Monte Carlo error in the analysis, a wide range of data streams were generated with different values of $\psi, \psi_1, \ldots, \psi_p$.

Tables \ref{tab:7} to \ref{tab:8} show results tracking $\alpha$-depth regions for $\alpha = 0.05$, 0.2 and 0.4 for periods $T=10^3$ and $T=10^4$ under optimal choices of the tuning parameter\footnote{In a practical situation, the history of the data stream can be used to estimate (or track) optimal values of the tuning parameters. We are currently working on such procedures.}. More detailed results are given in Figures 11 and 12 in supplementary material S.2. For $T = 10^3$ MADE is around 0.05
%This is a noticeable error, but the algorithm still can be useful for data streams with such rapid dynamics.
and estimation error does not decrease with increasing number of directional vectors which may seem surprising. The reason is that if the quantile estimates are poor, the intersections of the resulting halfspaces in \eqref{eq:11}, do not necessarily become better by adding more halfspaces.
%One poor halfspace estimate is enough to ruin the envelope estimate and by increasing the number of halfspace estimates, the chance of this happening will increase. 
For $T=10^4$ MADE is between 0.02 and 0.03. The optimal number of halfsspaces increases with dimension, but not dramatically.
%For period $T = 10^3$ the optimal number of halfspaces for $p=2$ and $p=5$ are 25 and 50, respectively and for $T = 10^4$, the optimal number of halfspaces are 25 and 500, respectively. Further, a higher number of halfspaces are optimal for a period $T = 10^4$ compared to a period $T = 10^3$ and is as expected. When the estimates of the halfspaces become better, more halfspaces can be used without getting into trouble that the intersections in Equation \eqref{eq:11} underestimate the sizes of the contours. 
\begin{table}
  \centering
  \begin{tabular}{lcccccccc}
    \hline
        & \multicolumn{2}{c}{$p=2$} & \multicolumn{2}{c}{$p=3$} & \multicolumn{2}{c}{$p=4$} & \multicolumn{2}{c}{$p=5$} \\\hline
    $n_u$ & MADE & Freq & MADE & Freq & MADE & Freq & MADE & Freq \\\hline
5    & 0.0559 &   972.7 &     $-$  &     $-$ &     $-$  &     $-$ &     $-$  &     $-$\\
10   & 0.0475 &   478.9 & 0.0577  &  486.7 &     $-$  &     $-$ &     $-$  &     $-$\\
25   & \textbf{0.0445} &   189.2 & \textbf{0.0457}  &  189.7 & 0.0510  &  184.7 &     $-$  &     $-$\\
50   & 0.0474 &    95.2 & 0.0467  &   95.1 & \textbf{0.0492}  &   93.3 & \textbf{0.0521}  &   92.4\\
100  & 0.0504 &    47.9 & 0.0502  &   47.4 & 0.0514  &   46.8 & 0.0523  &   46.4\\
200  &     $-$ &      $-$ & 0.0536  &   23.4 & 0.0546  &   22.6 & 0.0541  &   22.7\\
500  &     $-$ &      $-$ &     $-$  &     $-$ & 0.0590  &    9.1 & 0.0576  &    8.9\\
1000 &     $-$ &      $-$ &     $-$  &     $-$ &     $-$  &     $-$ & 0.0604  &    4.5\\\hline
  \end{tabular}
  \caption{Tracking of $\alpha$-depth regions for $\alpha = 0.05$, 0.2 and 0.4 for the distribution characterized by \eqref{eq:18} and \eqref{eq:19} with a period $T = 10^3$. The columns 'Freq' refers to how many times per millisecond the algorithm can update an $\alpha$-depth region when running on a single 1.8 GHz CPU processor.}
  \label{tab:7}
\end{table}
\begin{table}
  \centering
  \begin{tabular}{lcccccccc}
    \hline
        & \multicolumn{2}{c}{$p=2$} & \multicolumn{2}{c}{$p=3$} & \multicolumn{2}{c}{$p=4$} & \multicolumn{2}{c}{$p=5$} \\\hline
    $n_u$ & MADE & Freq & MADE & Freq & MADE & Freq & MADE & Freq \\\hline
5    & 0.0439 &   976.3 &     $-$   &    $-$ &     $-$  &     $-$ &     $-$  &     $-$\\
10   & 0.0305 &   480.1 & 0.0499   & 484.3 &     $-$  &     $-$ &     $-$  &     $-$\\
25   & \textbf{0.0226} &   189.2 & 0.0318   & 188.5 & 0.0429  &  184.7 &     $-$  &     $-$\\
50   & 0.0227 &    95.4 & 0.0277   &  94.3 & 0.0342  &   93.6 & 0.0395  &   91.1\\
100  & 0.0236 &    48.0 & \textbf{0.0275}   &  47.3 & 0.0306  &   47.0 & 0.0337  &   46.1\\
200  &     $-$ &      $-$ & 0.0289   &  23.3 & \textbf{0.0299}  &   22.7 & 0.0312  &   22.6\\
500  &     $-$ &      $-$ &     $-$   &    $-$ & 0.0307  &    9.1 & \textbf{0.0309}  &    9.0\\
1000 &     $-$ &      $-$ &     $-$   &    $-$ &     $-$  &     $-$ & 0.0316  &    4.5\\\hline
  \end{tabular}
  \caption{Tracking of $\alpha$-depth regions for $\alpha = 0.05$, 0.2 and 0.4 for the distribution characterized by \eqref{eq:18} and \eqref{eq:19}  with a period $T = 10^4$. The columns 'Freq' refers to how many times per millisecond the algorithm can update an $\alpha$-depth region when running on a single 1.8 GHz CPU processor.}
  \label{tab:8}
\end{table}

The algorithm is computationally very efficient. For dimension $p=5$ the algorithm can optimally process $10^4$ to $10^5$ observations from a data stream every second on a single CPU processor.
By using more equidistant directional vectors, we expect reduction in tracking error. Consider the dynamic case above except that the directional vectors are chosen more equidistantly. Directional vectors were generated using the filtering procedure in Section \ref{sec:alg} with $N_u = 10n_u$.
%As described in Section \ref{sec:alg}, this will increase the computational time in the initialization of the algorithm and thus the experiments were restricted to dimension $p=2$.

The results are shown in Table \ref{tab:9} and more detailed results are given in Figure 14 in supplementary material S.2.
\begin{table}
  \centering
  \begin{tabular}{lcc}
    \hline
$n_u$ & $T = 10^3$ & $T = 10^4$ \\\hline
5  & 0.0465& 0.0303\\
10 & \textbf{0.0400}& \textbf{0.0216}\\
25 & 0.0440& 0.0217\\
50 & 0.0477& 0.0226\\
100& 0.0505& 0.0236\\\hline    
  \end{tabular}
  \caption{Tracking of $\alpha$-depth regions for $\alpha = 0.05$, 0.2 and 0.4 for the distribution characterized by \eqref{eq:18} and \eqref{eq:19} using fairly equidistant directional vectors. Tracking error is measured using MADE. The left and right columns show results for $T = 10^3$ and $10^4$, respectively. Dimension is $p=2$.}
  \label{tab:9}
\end{table}
By comparing Tables \ref{tab:7} and \ref{tab:8} with \ref{tab:9}, we see that for $T = 10^3$ and $T = 10^4$, minimum MADE is reduced from 0.045 to 0.040 and from 0.0226 to 0.0216, respectively. However more importantly, by using equidistant vectors, the best results are obtained using fewer directional vectors. For both $T = 10^3$ and $T = 10^4$, the optimal number of vectors are reduced from 25 to 10. Finally we observe significant improvement if only five directional vectors were used. Using equidistant directional vectors adds an additional computational cost in the initialization of the algorithm, but will result in gained peak performance and fewer directional vectors, and thus less computation time and memory, needed during tracking. 

\section{Real-life Data Examples}
\label{sec:real}

In this section we use the algorithm on a real-life dataset related to activity change detection. A second real-life data example related to real-time event detection using Twitter data is given in supplementary material S.3.

We demonstrate how the algorithm can be used to detect outliers and events and perform classifications in dynamic settings. For example related to event detection, by characterizing a data stream distribution with multiple depth contours, in practice \textit{any} change in the data stream distribution can be detected. Not only changes in common properties such as expectation and covariance structure, but also changes in shape such as a change from an elliptic to a non-elliptic distribution. 

\subsection{Activity Change Detection}

Activity recognition is a highly active field of research where sensory information is used to automatically detect and identify activities of users. E.g. to detect sedentary lifestyle and prompt the user to perform healthy exercises. We will focus on identifying changes in activities using observations from accelerometer which is available on almost any smart cell phone today. 

We consider an accelerometer dataset from the WISDM (Wireless Sensor Data Mining) project \citep{kwapisz2011activity}. Accelerations in $x$, $y$ and $z$ directions where observed, with a frequency of 20 observations per second, while users were performing the activities walking, jogging, walking up a stairway and walking down a stairway. A total of 36 users were observed and the dataset contains a total of $989\,875$ observations. 

Current research focuses on supervised approaches where historic and annotated activity observations are used to train an activity classification model. E.g. \cite{kwapisz2011activity} trained models such as decision trees and neural networks. However such an approach is highly sensitive to any temporal changes in the data, e.g. if the user changes to an activity that is not part of the training material, becomes fitter, sick etc. In this example we rater take an unsupervised approach and the goal is to detect whenever the user changes activity. Since we receive 20 accelerometer observations per second, it is important that the streaming approach is computationally efficient.

Change detection is useful as part of a supervised scheme. Whenever a change is detected, the observations from the last activity can be classified and the supervised classifier retrained. If the supervised learner is sufficiently uncertain about the last activity, it may in real time ask the user for feedback.

\begin{figure}
  \centering
    \includegraphics[width = \textwidth]{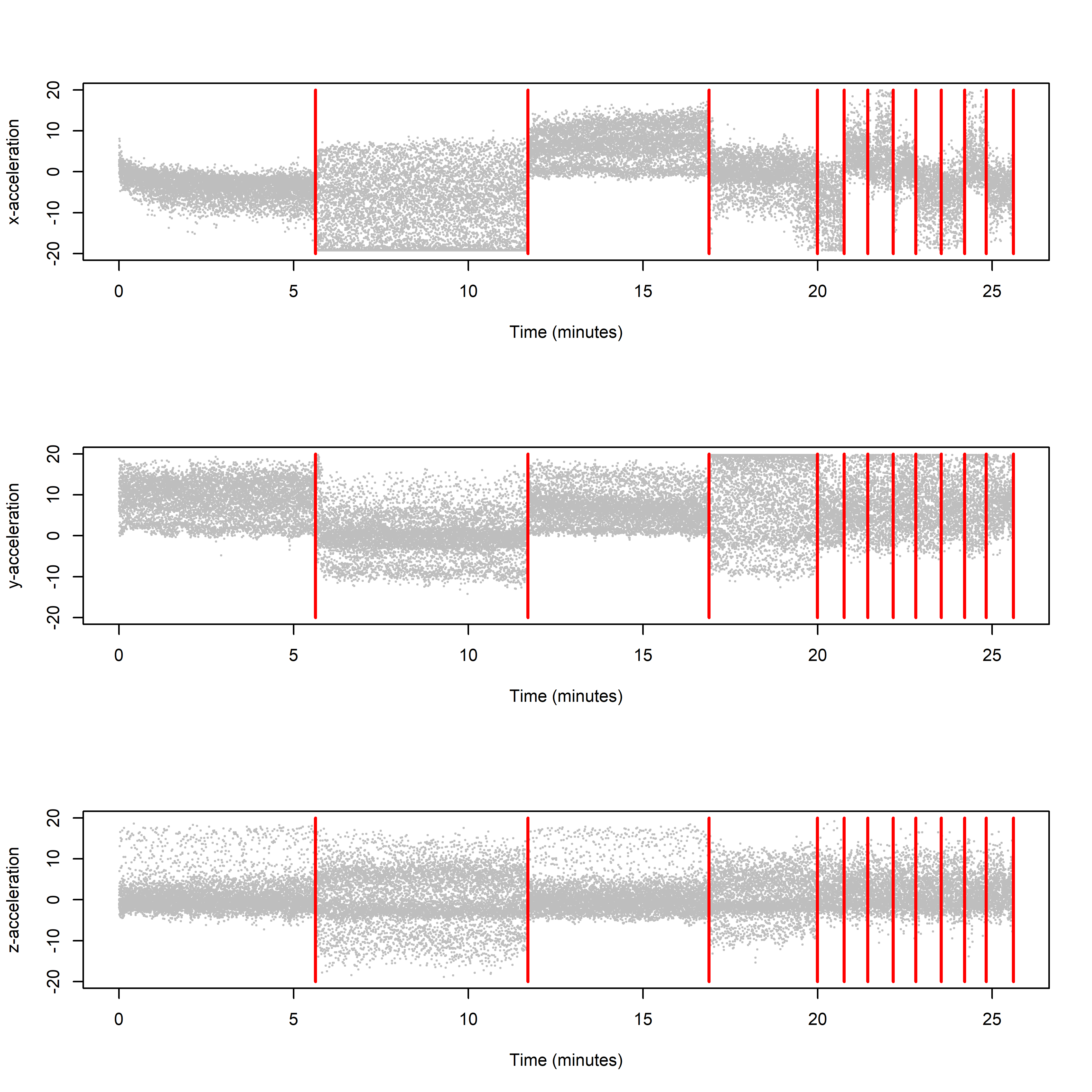}
    \caption{The gray dots show accelerometer observations for an arbitrary users. The red lines show when the user changes activity.}
  \label{fig:22}
\end{figure}

Figure \ref{fig:22} shows in gray $x$, $y$ and $z$ acceleration for an arbitrary user. The red lines show when the user changed activity. Acceleration distributions are fairly stationary within an activity, but with some gradual and abrupt changes. The users often changed activities as often as every 30 second making this a challenging tracking and change detection problem. 

\begin{figure}
  \centering
  \begin{tabular}{cc}
    \includegraphics[width = 0.5\textwidth]{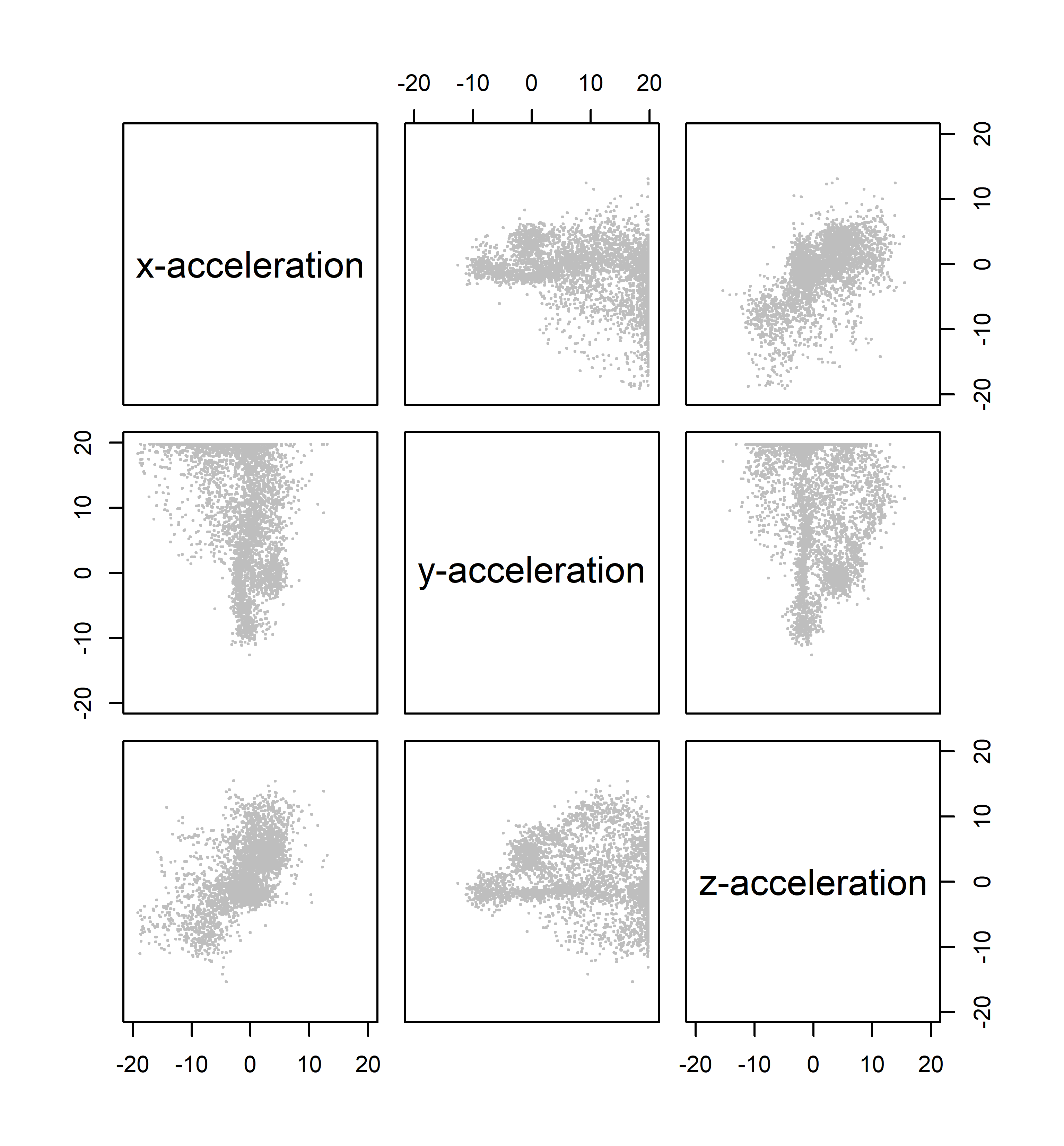} & \includegraphics[width = 0.5\textwidth]{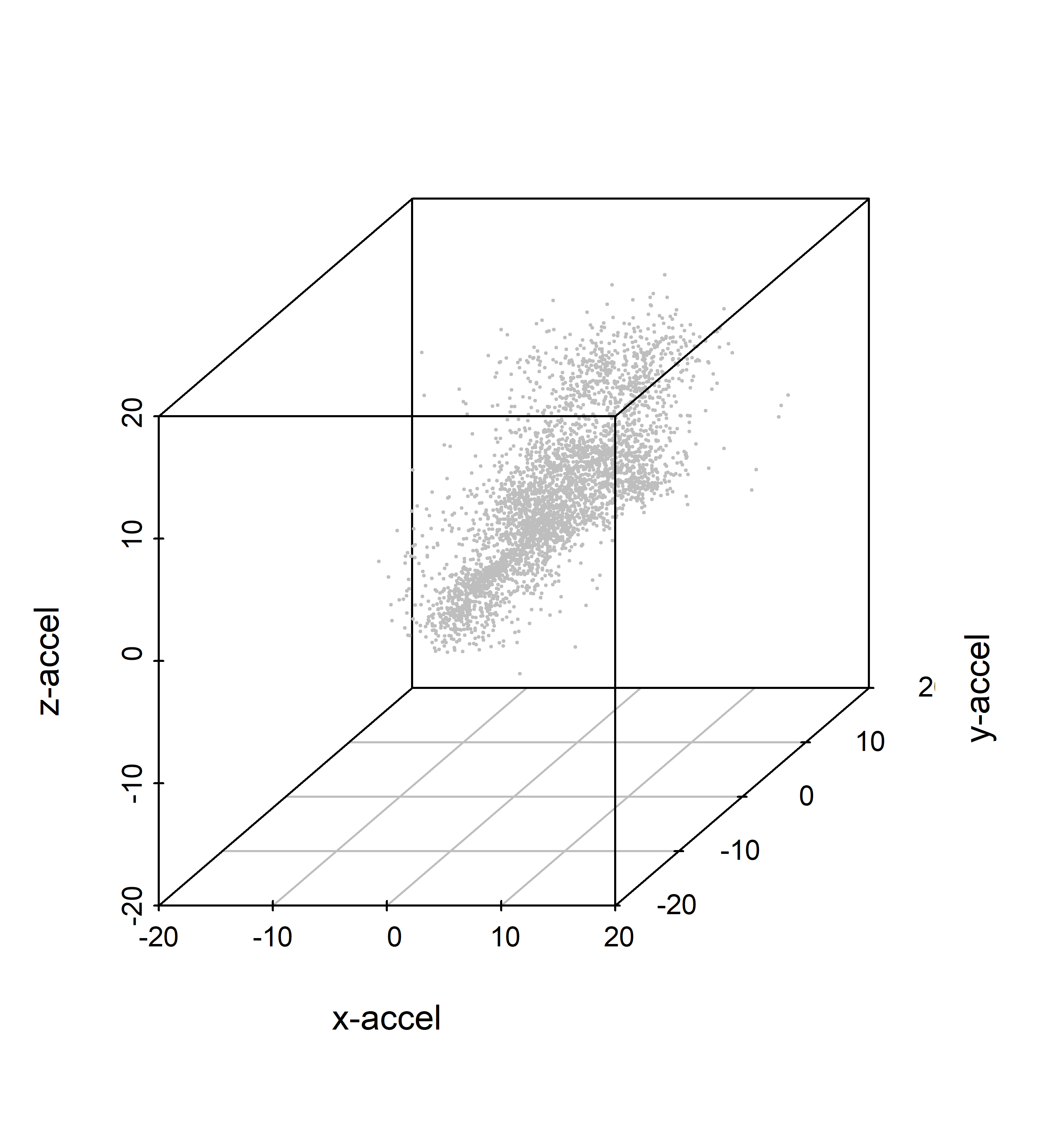}\\
    \includegraphics[width = 0.5\textwidth]{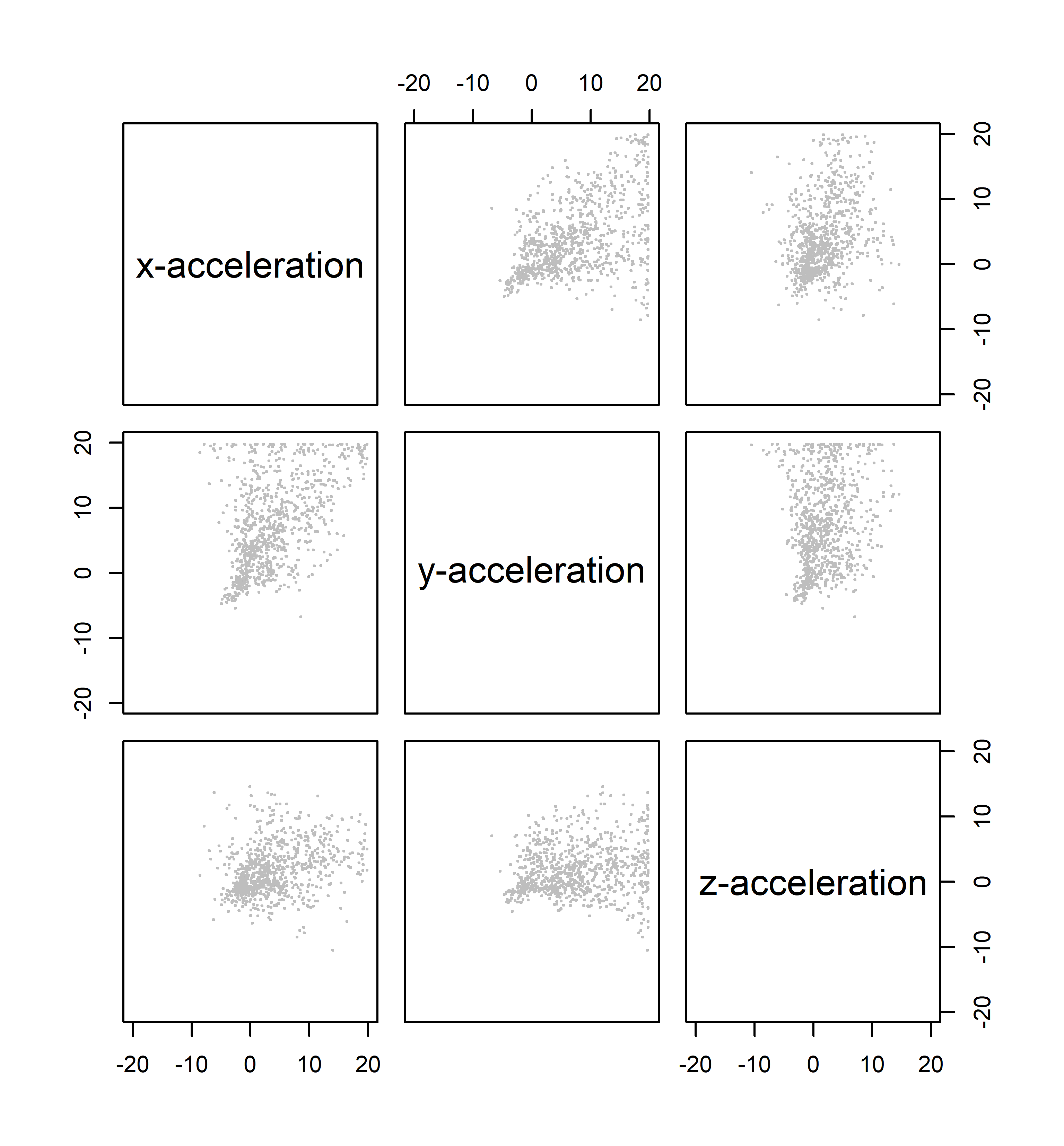} & \includegraphics[width = 0.5\textwidth]{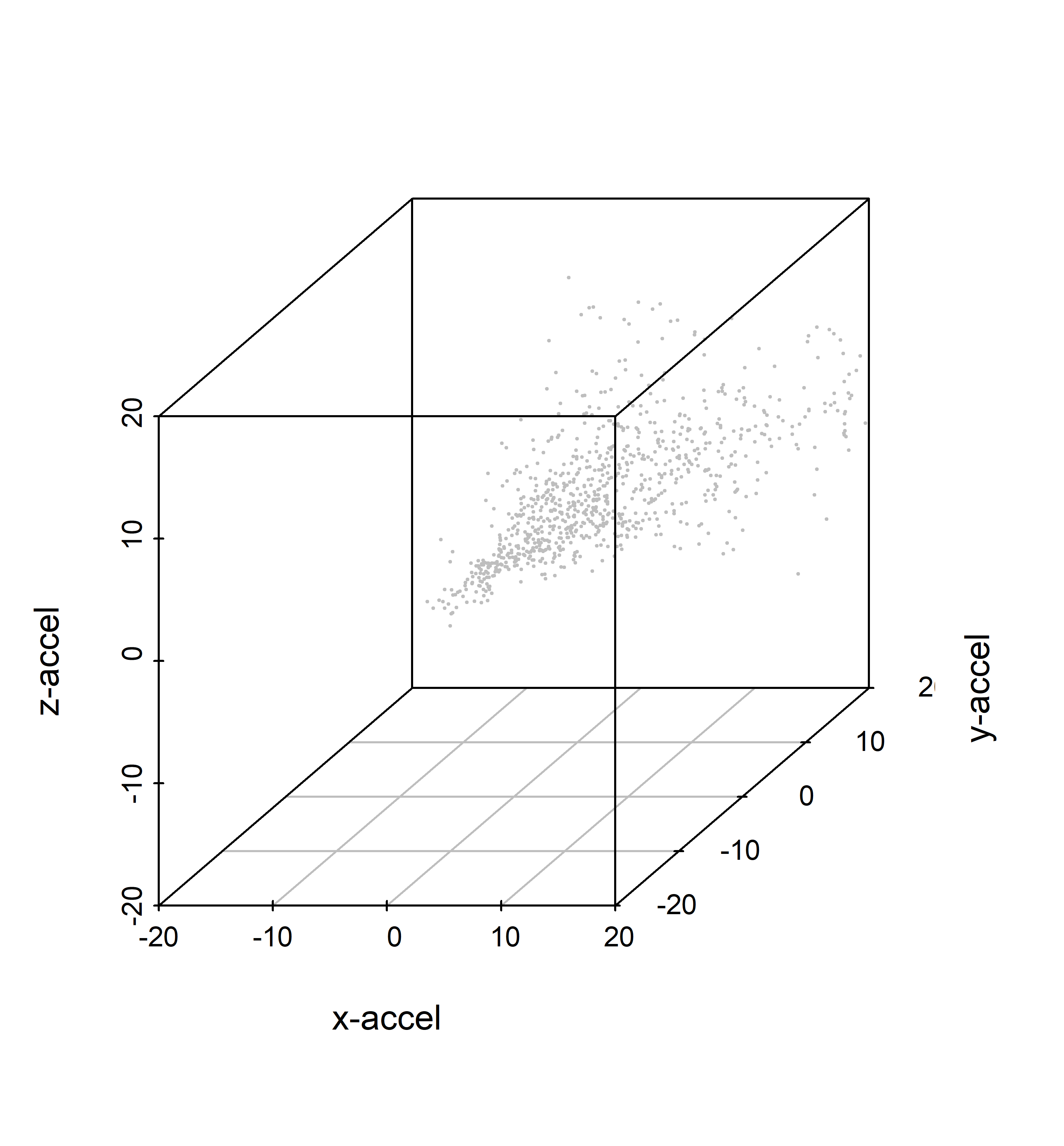}
  \end{tabular}
    \caption{The first and the second row show scatterplot of accelerometer observations for two activity sessions.}
  \label{fig:23}
\end{figure}
Figure \ref{fig:23} shows scatterplots of the accelerometer observations for two arbitrary sessions with minimal temporal trend. The simultaneous acceleration distributions vary a lot between sessions and are often far from elliptical. Further, even though the distributions may look quite different, the mean and covariances often are quite similar making change detection based on elliptic distributions challenging. We thus suggest the following simple depth based change detection procedure:
\begin{enumerate}
\item Track $\alpha$-depth contours of the simultaneous acceleration distribution by tracking $n_u$ directional quantiles using the DUMIQE algorithm with tuning parameter $\lambda$.%\footnote{We could also have used the ShiftQ algorithm, but since joint estimates is not important for this application, we used the DUMIQE algorithm.}
\item Compute the Euclidean distance between the current $\alpha$-depth contours and the contours $h$ seconds back in time using Equation \eqref{eq:16}. Let ED$_t$ denote the distance at time $t$.
\item Track the expectation and standard deviation of ED$_t$ distribution using exponential moving average
  \begin{align*}
    E(\text{ED}_t) &= (1 - \delta) E(\text{ED}_{t-1}) + \delta \text{ED}_t \\
    E(\text{ED}_t^2) &= (1 - \delta) E(\text{ED}_{t-1}^2) + \delta \text{ED}_t^2 \\
    SD(\text{ED}_t) &= \sqrt{E(\text{ED}_t)^2 - E(\text{ED}_t^2)}
  \end{align*}
\item When the user changes activity, we expect ED$_t$ to rapidly increase. We detected a new activity when ED$_t$ is more than $\eta$ standard deviations higher then $E(\text{ED}_t)$, i.e. $\text{ED}_t \geq E(\text{ED}_t) + \eta \, SD(\text{ED}_t)$.
\item When a new activity was detected, restart the tracking of the $\alpha$-depth contours and go back to step 1. 
\end{enumerate}
The beauty of the approach above is that since it measures difference in depth contours, it can in practice detect \textit{any} kind of changes in the shape simultaneous acceleration distribution: for example a change from a symmetric to a non-symmetric distribution. Given the properties of the observations in this application, this flexibility is important.

We compare the approach against an identical approach except that in the first part of the algorithm the mean and covariance structure (and not depth contours) were tracked using multivariate exponentially weighted moving average (MEWMA) \citep{lowry1992multivariate}. 
%of the simultaneous acceleration distribution. The approach thus is similar to the well-known multivariate exponentially weighted moving average control chart (MEWMA) \citep{lowry1992multivariate} and goes as follows:
%\begin{enumerate}
%\item Let $(x_{1t}, x_{2t}, x_{3t})$ denote the observed $x, y$ and $z$ accelerations at time $t$. Track the mean and covariances of the simultaneous distribution using exponential moving average
%  \begin{align*}
%    {\text{E}}(X_{it}) &= (1 - \nu) {\text{E}}(X_{it}) + \nu x_{it},\,\, i = 1,2,3 \\
%    {\text{E}}(X_{it} X_{jt}) &= (1 - \nu) {\text{E}}(X_{it} X_{jt}) + \nu x_{it}x_{jt},\,\, i,j = 1,2,3 \\
%    {\Sigma}_{ijt} &= {\text{E}}(X_{it} X_{jt}) - {\text{E}}(X_{it}) {\text{E}}(X_{jt}),\,\, i,j = 1,2,3
%  \end{align*}
%\item Compute the Mahalanobis distance between the current estimate of the mean vector and $h$ seconds back in time, $MD_t = (\mu_t - \mu_{t-h})^T {\Sigma}_t^{-1} (\mu_t - \mu_{t-h})$.
%\item Track the expectation and standard deviation of MD$_t$ distribution using exponential moving average in the same way as for ED$_t$ in the procedure above.
%\item When the user changes activity, we expect MD$_t$ to rapidly increase. We detected a new activity when MD$_t$ where more than $\eta$ standard deviations higher then $E(\text{MD}_t)$, i.e. $\text{MD}_t \geq %E(\text{MD}_t) + \eta \, SD(\text{MD}_t)$.
%\item When a new activity is detected, restart the tracking of the mean vector and covariance matrix and go back to step 1. 
%\end{enumerate}

We measured the performance of the depth and the MEWMA approaches for a wide range of values for the tuning parameters. Several sessions lasted for only 30 seconds and it was thus important for the tracking algorithms to rapidly adapt to a session before a new change of activity took place. In the first step of the procedures we thus chose decreasing values of the tuning parameters, but with a minimum value to take into account the dynamic changes in accelerations within a session, $\lambda_t = \max\{1/t, \lambda_{\text{min}}\}$, and tried the values 0.1, 0.05 and 0.01 for $\lambda_{\text{min}}$\footnote{For MEWMA, $\lambda_t$ refers to the moving average tuning parameter and $\lambda_t = 1/t$ is thus equivalent to the sample mean.}. This performed better than using constant values of the tuning parameter. We further tried the values 0.1, 0.05 and 0.01 for $\delta$, 2.5, 5 and 10 seconds for $h$ and 2, 5 and 8 for $\eta$. Further for the depth approach we used three depth contours with $\alpha$ equal to 0.2, 0.05 and 0.01 and tried $n_u = 20$ or 50 directional vectors. We ran the two change detection approaches for the whole dataset for all the combinations of the parameters. This resulted in a total of 162 and 81 experiments for the depth and the MEWMA approaches, respectively. 

Precision, recall and the F1 score was used to measure performance \citep{sokolova2009systematic}. If the approach detects more than one change between two true changes, we characterize the first change as a correct detection and the others as false detections and define
\begin{align*}
  \text{Precision} &= \frac{ \text{No. of correct detections} }{ \text{No. of detections} } \\[2mm]
  \text{Recall} &= \frac{ \text{No. of correct detections} }{ \text{No. true changes} } \\[2mm]
  \text{F1 score} &= \frac{ 2\,\cdot\text{Precision}\,\cdot\, \text{Recall}  }{\text{Precision} + \text{Recall}} 
\end{align*}
%where the F1 score is the harmonic mean of precision and recall.

Tables \ref{tab:2} and \ref{tab:3} show the top ten results with respect to the F1 score. The depth approach outperforms the MEWMA with respect to the F1 score and in addition detects the true changes more rapidly.
%For the depth approach, the overall best results are achieved when the precision is high relative to the recall, while for the MEWMA the recall is higher than the precision. However, there are exceptions, e.g. for the third and the sixth rows of the depth approach results, the recalls are higher than the precisions and still the overall F1 scores are higher than for all the MEWMA results.
The performance of the depth approach does not seem to be particularly sensitive on the number of directional quantiles used.
%For both approaches the best results are achieved using a high value for the detection threshold $\eta$. 
\begin{table}
  \centering
  \begin{tabular}{ccccc|cccc}
  \hline
   $\lambda_{\text{min}}$ & $\delta_{\text{min}}$ & $h$ & $\eta$ & $n_u$ & Precision & Recall & F1 score & Det. delay (sec) \\ \hline
 0.01 &  0.01 & 100 &  8 &  20 &   0.796  & 0.497  & \textbf{0.612}  & \textbf{1.325}\\
 0.01 &  0.01 & 100 &  8 &  50 &   0.781  & 0.486  & \textbf{0.599}  & \textbf{1.150}\\
 0.01 &  0.05 & 200 &  8 &  20 &   0.532  & 0.682  & \textbf{0.597}  & \textbf{1.415}\\
 0.01 &  0.05 &  50 &  8 &  50 &   0.613  & 0.564  & \textbf{0.587}  & \textbf{1.876}\\
 0.01 &  0.01 & 200 &  8 &  20 &   0.735  & 0.488  & \textbf{0.587}  & \textbf{1.202}\\
 0.01 &  0.05 & 200 &  8 &  50 &   0.510  & 0.673  & \textbf{0.580}  & \textbf{1.482}\\
 0.01 &  0.01 & 200 &  8 &  50 &   0.719  & 0.480  & \textbf{0.575}  & \textbf{1.219}\\
 0.01 &  0.05 & 100 &  8 &  20 &   0.538  & 0.618  & \textbf{0.575}  & \textbf{1.010}\\
 0.05 &  0.10 & 200 &  8 &  20 &   0.539  & 0.616  & \textbf{0.575}  & \textbf{1.903}\\
 0.01 &  0.05 &  50 &  8 &  20 &   0.613  & 0.532  & \textbf{0.570}  & \textbf{1.894}\\\hline
  \end{tabular}
  \caption{Change detection example. Results for the depth approach.}
  \label{tab:2}
\end{table}
\begin{table}
  \centering
  \begin{tabular}{cccc|cccc}
  \hline
   $\lambda_{\text{min}}$ & $\delta_{\text{min}}$ & $h$ & $\eta$ & Precision & Recall & F1 score & Det. delay (sec) \\ \hline
 0.01  &   0.01 &   200  &    8  & 0.454  & 0.697  & \textbf{0.550}  & \textbf{1.553}\\
 0.05  &   0.01 &   200  &    8  & 0.447  & 0.697  & \textbf{0.545}  & \textbf{1.691}\\
 0.05  &   0.01 &    50  &    8  & 0.438  & 0.699  & \textbf{0.539}  & \textbf{2.249}\\
 0.01  &   0.01 &    50  &    8  & 0.421  & 0.711  & \textbf{0.529}  & \textbf{1.736}\\
 0.01  &   0.01 &   100  &    8  & 0.398  & 0.737  & \textbf{0.517}  & \textbf{1.293}\\
 0.05  &   0.01 &   100  &    8  & 0.388  & 0.711  & \textbf{0.502}  & \textbf{1.747}\\
 0.05  &   0.05 &   200  &    8  & 0.353  & 0.760  & \textbf{0.483}  & \textbf{1.525}\\
 0.05  &   0.05 &    50  &    8  & 0.336  & 0.818  & \textbf{0.476}  & \textbf{1.832}\\
 0.05  &   0.10 &   200  &    8  & 0.336  & 0.803  & \textbf{0.474}  & \textbf{1.522}\\
 0.01  &   0.05 &   200  &    8  & 0.330  & 0.777  & \textbf{0.463}  & \textbf{1.281}\\\hline
  \end{tabular}
  \caption{Change detection example. Results for the MEWMA approach.}
  \label{tab:3}
\end{table}
%D the properties of accelerometer distributions in the example, the depth tracking approach ourperforms MED
%depth tracking approach perform well since the acceleration distributions often are far from elliptical
% will perform better than using a tracking approach based on elliptic distribution assumptions. 

%\subsection{Robust PCA}

%Done some experiments on intuition data. Not easy to interpret the PCAs

%\subsection{Detect forged banknotes}

%Online clasification. Depth not very useful. Eliptical fit seems to perform well here.

\section{Closing Remarks}

In this paper we have presented a computationally and memory efficient procedure to estimate and track Tukey $\alpha$-depth contours using incremental quantile estimators. Due to the flexibility of Tukey depth, the procedure characterize elliptic and non-elliptic distributions equally well. The real-life data examples demonstrate that the procedure is useful to make real-time decisions from complex multidimensional streaming data.

To estimate $\alpha$-depth contours, the number of directional vectors, $n_u$, and values of tuning parameters in the incremental quantile tracking algorithms must be chosen. We are currently working on procedures that use use information from the history of the data stream to recursively update such values. 

\section{Supplementary Material}

Download from here: \url{https://www.dropbox.com/s/p3uoayg411dzs2k/DepthArXiVSuplMaterial.pdf?dl=0}

\bibliographystyle{dcu}
\bibliography{bibl}

\begin{appendices}
\setcounter{section}{1}
  
\section*{Appendices}

\subsection{Computation of Depth Error for Elliptic Distributions}
\label{app:de}

For elliptic distributions, the $\alpha$-depth contours coincide with the contours of the distribution and, secondly, the Tukey halfplanes are tangent planes to the contours \citep{kong2012quantile}. For a multivariate normal distribution with expectation vector $\mu$ and covariance matrix $\Sigma$, the depth of any point $w$ can be found analytically. First find the inward pointing unit length normal vector to the tangent plane at $w$:
  \begin{align*}
    \text{tang}\,(w) = - \frac{\Sigma^{-1}(w - \mu)}{\|\Sigma^{-1}(w - \mu)\|_2}
  \end{align*}
and then compute the depth as the probability to be within the tangent plane halfspace defined by $\text{tang}\,(w)$
  \begin{align*}
    D(w,P) = \Phi\left(\text{tang}\,(w)^Tw; \text{tang}\,(w)^T\mu, \sqrt{\text{tang}\,(w)^T \Sigma^{-1} \text{tang}\,(w)}\right)
  \end{align*}
  where $\Phi(\,\cdot\,; m, s)$ refers to the cumulative distribution function of the univariate normal distribution with expectation $m$ and standard deviation $s$.
  % This approach makes it possible to evaluate the performance of our approach for high dimensions $p$ and thus will be important in our experiments.

\end{appendices}

\end{document}